\documentclass[11pt, a4paper, logo, copyright]{googledeepmind}

\ifdefined\pdfinfoomitdate
\fi
\ifdefined\pdftrailerid
  \pdftrailerid{redacted}
\fi

\makeatletter
\providecommand\bibentry[1]{\nocite{#1}\fullcite{#1}}
\makeatother

\usepackage{kantlipsum, lipsum}
\usepackage{dsfont}
\usepackage{gdm-colors}
\usepackage[utf8]{inputenc}   \usepackage{newunicodechar}   \usepackage{amssymb}          \usepackage{CJKutf8}
\usepackage{tcolorbox}
\tcbuselibrary{breakable}
\usepackage{enumitem}
\usepackage{longtable}      \usepackage{arydshln}
\usepackage[flushleft]{threeparttable}
\usepackage{wasysym,marvosym}

\definecolor{catBlue}{HTML}{2A6FB5}
\definecolor{catPurple}{HTML}{6E3FA3}
\definecolor{catTeal}{HTML}{1F6F6B}
\definecolor{catRed}{HTML}{C0392B}
\definecolor{cellGreen}{HTML}{6CC180}
\definecolor{cellGray}{HTML}{ECECEC}

\usepackage{graphicx}

\usepackage{caption}
\usepackage{xcolor}
\usepackage{soul}

\soulregister\cite7
\soulregister\ref7
\soulregister\cref7

\usepackage[
  style=authoryear-comp,
  sorting=ynt,
  natbib=true,
  backend=bibtex,
  maxcitenames=2,
  maxbibnames=99,
  uniquelist=false,
  uniquename=false,
  giveninits=true,
  dashed=false
]{biblatex}
\let\cite\textcite

\DeclareFieldFormat{citehyperref}{%
  \DeclareFieldAlias{bibhyperref}{noformat}%
  \bibhyperref{#1}}
\DeclareFieldFormat{textcitehyperref}{%
  \DeclareFieldAlias{bibhyperref}{noformat}%
  \bibhyperref{%
    #1%
    \ifbool{cbx:parens}
      {\bibcloseparen\global\boolfalse{cbx:parens}}%
      {}}}
\savebibmacro{cite}
\savebibmacro{textcite}
\renewbibmacro*{cite}{%
  \printtext[citehyperref]{%
    \restorebibmacro{cite}%
    \usebibmacro{cite}}}
\renewbibmacro*{textcite}{%
  \ifboolexpr{
    ( not test {\iffieldundef{prenote}} and
      test {\ifnumequal{\value{citecount}}{1}} )
    or
    ( not test {\iffieldundef{postnote}} and
      test {\ifnumequal{\value{citecount}}{\value{citetotal}}} )
  }%
    {\DeclareFieldAlias{textcitehyperref}{noformat}}%
    {}%
  \printtext[textcitehyperref]{%
    \restorebibmacro{textcite}%
    \usebibmacro{textcite}}}

\usepackage{bbding}
\usepackage[utf8]{inputenc} \usepackage[T1]{fontenc}    \usepackage{hyperref}       \usepackage{url}            \usepackage{booktabs}       \usepackage{nicefrac}       \usepackage{microtype}      \usepackage{amsmath}
\usepackage{graphicx}
\usepackage{tablefootnote}
\usepackage{multicol}
\usepackage[nameinlink]{cleveref}
\crefname{figure}{Figure}{Figures}
\Crefname{figure}{Figure}{Figures}
\crefname{table}{Table}{Tables}
\Crefname{table}{Table}{Tables}
\usepackage{bbm}
\usepackage{multirow}
\usepackage{soul}
\usepackage{float}
\usepackage{placeins}
\usepackage{wrapfig}
\usepackage{blindtext}
\usepackage{tablefootnote}
\usepackage{amsfonts}
\usepackage{colortbl}
\usepackage{mathtools,amssymb}
\usepackage{bm}
\usepackage{makecell}
\usepackage{caption}
\usepackage{capt-of}
\usepackage{array}
\usepackage{calc}      \usepackage{caption}   \usepackage{subcaption}  \usepackage{xcolor,colortbl}\usepackage{tikz}
\usepackage[bottom]{footmisc}

\usepackage{xcolor}   \usepackage{soul}
\sethlcolor{yellow!20}

\definecolor{mydeepgreen}{RGB}{0, 100, 0}

\usepackage{xspace}

\newcommand{\cdashlinerow}[2]{  \cdashline{#1}  \noalign{\global\let\CT@row@color\relax\vskip0pt}  \rowcolor{#2}}

\newcommand{\eat}[1]{}

\crefformat{section}{\S#2#1#3}

\graphicspath{{figures/}}

\title{X-Rec Technical Report}

\renewcommand{\today}{}

\author{\sffamily\bfseries\large TikTok-Data-Content Intelligence \& TikTok-Data-Feed Quality}

\begin{abstract}
Recent advances in generative modeling have reshaped recommender systems by formulating recommendation as a next-item generation problem. Existing retrieval approaches primarily follow two paradigms: user-to-item (U2I) methods represent user context using one or a few deterministic embeddings, which limits the ability to capture diverse and multi-mode interests, while semantic-ID-based autoregressive (SID-AR) methods model more expressive distributions but suffer from quantization errors and the low throughput of sequential decoding. To address these limitations, we propose X-Rec to directly learn the recommendation distribution in the continuous item embedding space through flow matching and generate embedding triggers for approximate nearest neighbor retrieval. X-Rec incorporates three key designs to make this formulation effective and efficient. First, we introduce anchor conditioning to decompose generation into coarse semantic-region selection and fine-grained refinement. Second, we adopt Riemannian flow matching to align generative trajectories with the hyperspherical geometry of item embeddings. Third, we design a late-interaction diffusion Transformer that restricts repeated velocity-field estimation to the final Transformer layer. On a streaming benchmark, X-Rec substantially outperforms U2I baselines, matches the retrieval quality of SID-AR methods, and delivers \(3.46\times\) higher inference throughput than SID-AR. X-Rec has also been deployed as a new retrieval source for a specific vertical content on TikTok, where two consecutive launches have yielded significant improvements in both vertical engagement (+4.1484\%) and general engagement (+0.0111\%).

\end{abstract}

\begin{document}

\maketitle

\newpage
\setcounter{tocdepth}{2}
\tableofcontents

\newpage
\section{Introduction}

Recommendation systems have become a core infrastructure for modern content platforms, connecting users with videos, services, merchants, and local consumption opportunities on an industrial scale~\citep{ricci2010introduction,ricci2021recommender}. In large-scale retrieval, a central challenge is to efficiently recall compact and diverse items that reflects a user's evolving interests before downstream ranking.

 Recommendation can be formulated as a next-item generation task that models the recommendation distribution \(p(x \mid c)\), where \(x\) denotes the next item and the user context \(c\) typically comprises the user's historical interaction sequence \(s\) and the expected target attribute \(f\)\footnote{For example, \(f\) can denote the expected target action to be modeled.}~\citep{covington2016deep,kang2018self,rajput2023recommender}. To this end, existing approaches can be broadly categorized into two paradigms. The first is two-tower user-to-item (U2I) retrieval~\citep{cen2020controllable,chen2024hllm,zhang2024notellm}, which represents each item \(x\) as a continuous embedding and encodes the user context \(c\) into one or a few fixed user embeddings for approximate nearest neighbor (ANN) retrieval. Although widely adopted in industrial systems~\citep{covington2016deep}, this paradigm is limited by its intrinsic \textit{poor expressiveness}: U2I models approximate the target distribution \(p(x \mid c)\) with a delta distribution. Such a distribution is too simple to capture diverse user interests intuitively and has limited model capacity theoretically, e.g., in terms of VC dimension~\citep{csikos2019tight}. Although recent multi-interest U2I approaches alleviate this problem by introducing more embeddings to encode \(c\), the underlying distributional formulation remains unchanged. Another line of work is SID-AR~\citep{rajput2023recommender,deng2025onerec,zhou2025onerec}, which discretizes item embeddings into semantic identifiers (SIDs), thereby transforming recommendation into an autoregressive next-token prediction problem. This approach defines tractable distributions over SIDs and can represent substantially more flexible distributions than the delta distribution. However, it also faces several practical and modeling challenges. The first challenge is what we refer to as \textit{flawed SIDs}, which manifests in two aspects: 1) Discretizing continuous item representations inevitably introduces quantization error~\citep{lee2022autoregressive}, leading to information loss in the item representation space; and 2) SIDs are ideally expected to preserve a near one-to-one correspondence with item IDs, which is difficult to satisfy in practice.\footnote{A common workaround is to append a non-semantic hash token to the SID sequence~\citep{rajput2023recommender,fu2026forge}, but this weakens the semantic structure of the code itself.} The second challenge is \textit{low throughput}: autoregressive models generate SID tokens sequentially, limiting inference throughput and posing a significant bottleneck for large-scale retrieval scenarios~\citep{stern2018blockwise,kwon2023pagedattention}.

In this paper, we represent each item by its embedding and directly model the conditional recommendation distribution \(p(x \mid c)\) using flow matching (FM)-based diffusion~\citep{lipman2022flow,liu2022flow}. We term this approach \textbf{X-Rec}, emphasizing a return to \(x\) itself as the fundamental object of recommendation. This design directly addresses the limitations of existing paradigms. First, to overcome the poor expressiveness of U2I, X-Rec treats recommendation as universal distribution estimation in the continuous item embedding space. Utilizing FM, it can capture complex, diverse, and multi-mode user interest distributions, rather than collapsing them into one or a few deterministic retrieval points. Second, to overcome flawed SIDs of SID-AR, X-Rec operates directly on item embeddings, avoiding the quantization errors introduced by SID discretization, and eliminating the need to enforce a fragile one-to-one correspondence between SIDs and item IDs. Third, to overcome the low throughput of SID-AR, X-Rec achieves high inference throughput by generating a single item embedding through diffusion, rather than autoregressively generating multiple SID tokens, making it an efficient alternative for large-scale retrieval.

Built upon diffusion transformers (DiT)~\citep{peebles2023scalable,wu2025qwen}, X-Rec incorporates three key designs to make continuous generative retrieval both effective and efficient. First, we introduce anchor conditioning to enable coarse-to-fine generation. Before generating the target item embedding, X-Rec first predicts an anchor, i.e., the cluster index of the target embedding, and then conditions the embedding generation process on this anchor. This design reduces the difficulty of directly generating fine-grained item embeddings from scratch, while also making the generation process more controllable. In particular, by manipulating the anchor distribution, X-Rec can explicitly encourage broader coverage over different regions of the item space, thereby improving the diversity of retrieved candidates. Empirically, this anchor-guided generation improves \(\operatorname{Recall}@20\) by 2.05\% percentage points, demonstrating its effectiveness in enhancing retrieval quality. Second, we adopt Riemannian flow matching (RFM)~\citep{chen2024flowgeometry,kumar2026learning}, where the generative trajectory evolves along geodesics on the hypersphere. This formulation is well aligned with the geometry of multi-mode item embeddings, which are typically learned through contrastive objectives under cosine similarity and thus naturally lie on a spherical manifold~\citep{radford2021learning}. Compared with rectified FM in Euclidean space, RFM avoids wasting model capacity on forcing generated embeddings back onto the sphere, and instead directly models the intrinsic geometry of the item embedding space. Experiments show that this geometry-aware formulation improves \(\operatorname{Recall}@20\) by 1.78\% in absolute, highlighting the benefit of modeling item embeddings on their native manifold. Third, we employ a late-interaction architecture for efficient velocity-field estimation. Instead of applying the full Transformer at every diffusion step, X-Rec uses only the final Transformer layer to compute the velocity field. Empirically, compared to the full-layer denoising setup, the one-layer design reduces \(\operatorname{Recall}@20\) by only 1.05\% percentage points, while improving generation throughput by approximately \(8.28\times\). This makes X-Rec more suitable for large-scale retrieval scenarios where both recommendation quality and serving efficiency are critical.

In offline evaluation, X-Rec substantially outperforms strong baselines. Compared to U2I methods, it can capture the diverse, multi-mode structure of user interests more precisely, leading to significant recall performance improvements. Compared to SID-AR, it achieves comparable recall performance, while improving the inference throughput by \(3.46\times\), making it a more practical generative retrieval solution for large-scale industrial deployment. X-Rec has also been deployed as a new retrieval source for a specific vertical content on TikTok. The online A/B tests show that two consecutive launches deliver significant gains in both vertical engagement (+4.1484\%) and general engagement metrics (+0.0111\%).

In the remainder of this paper, we first examine different generative retrieval paradigms in \cref{sec:formulations}, describing their formulations, advantages and limitations. Next, we elaborate on the proposed X-Rec framework in \cref{sec:x-rec_framework}, including its training pipeline and model architecture. Then, we conduct offline experiments to validate the effectiveness of X-Rec and its key components in \cref{sec:offline-experiments}, followed by online evaluation results in \cref{sec:online-results}. Finally, we conclude this work and propose potential future directions in \cref{sec:conclusion}.

\section{Formulations of Different Paradigms}\label{sec:formulations}

Let \(\mathcal{X}\) denote the item set. Given a user context \(c\), our goal is to model the conditional target item distribution \(p_\theta(x \mid c)\) over \(\mathcal{X}\), where \(\theta\) denotes the model parameters. Following standard sequential recommendation, the context \(c\) contains a chronologically ordered historical interaction sequence \(s\)~\citep{kang2018self,rajput2023recommender}. In our formulation, it additionally contains expected target attributes\footnote{We do not distinguish between item features, e.g., item type, and user--item interaction features, e.g., the user's action on the item. Both are included as attributes.} \(f=(f^1,f^2,\cdots,f^F)\in\mathcal{F}\), where $f^i$ denotes a sub-attribute. The behavior sequence \(s\) consists of a series of interaction tuples, each containing an interacted item and its associated attributes:
\begin{equation}
    s=(i_1,i_2,\cdots,i_N), \quad i_n=(x_n,f_n)
\end{equation}
where \(f_n=(f_n^1,f_n^2,\cdots,f_n^F)\) denotes the attribute vector associated with item \(x_n\), for \(n\in\{1,\cdots,N\}\).

During inference, we generate multiple retrieval triggers \(\tilde{x}_{1:K}\) from \(p_\theta(x \mid c)\) and use them to retrieve candidate items from \(\mathcal{X}\). Depending on how the item \(x\) and the recommendation distribution \(p_\theta(x \mid c)\) are represented, existing retrieval methods can be instantiated under different paradigms. In the following, we compare these paradigms in terms of both modeling capability and computational complexity. For the complexity analysis, we assume that all methods are built on a Transformer backbone~\citep{vaswani2017attention} with \(L\) layers and hidden dimension \(d\), where the condition length is \(N\) and the number of generated retrieval triggers is \(K\). \cref{tab:paradigm_comparison} summarizes the formalization and computational complexity of different methods.

\begin{table}[htbp]
\centering
\begin{tabular}{lllll}
\toprule
\multicolumn{1}{c}{\multirow{2}{*}{Paradigms}} & \multicolumn{2}{c}{Formalization} & \multicolumn{2}{c}{Complexity} \\
\multicolumn{1}{c}{} & \multicolumn{1}{c}{$x$} & \multicolumn{1}{c}{$p(x|c)$} & \multicolumn{1}{c}{Time} & \multicolumn{1}{c}{Space} \\
\midrule
U2I & Embeddings & Train: Sampled Categorical; Test: Delta & $\mathcal{O}(N^2)$ & $\mathcal{O}(1)$ \\
SID-AR & SIDs & Discrete Categorical & $\mathcal{O}(N(N+KM))$ & $\mathcal{O}(LN)$ \\
X-Rec & Embeddings & Continuous & $\mathcal{O}(N(N+M))$ & $\mathcal{O}(N)$ \\
\bottomrule
\end{tabular}
\caption{Formalization and complexity of different generative retrieval paradigms.}
\label{tab:paradigm_comparison}
\end{table}

\subsection{U2I}

The U2I approach~\citep{chen2024hllm,cen2020controllable} uses an embedding $x^{(e)}$ to represent item $x$. During training, it is commonly optimized with a contrastive objective, such as InfoNCE~\citep{oord2018representation,radford2021learning}, which can be interpreted as modeling a sampled categorical distribution:
\begin{equation}
    p_\theta^{\text{u2i-train}}(x|c)=\frac{\exp\left(\frac{e_\theta(c)^\top x^{(e)}}{\tau}\right)}{\exp\left(\frac{e_\theta(c)^\top x^{(e)}}{\tau}\right)+\sum_{x'\in\mathcal{X}_{\text{neg}}} \exp\left(\frac{e_\theta(c)^\top {x'}^{(e)}}{\tau}\right)}
\end{equation}
Here, $\tau$ denotes the temperature hyperparameter, and $\mathcal{X}_{\text{neg}}$ is a collection of negative samples, usually constructed via random sampling strategies such as in-batch negatives~\citep{chen2020simple}. Because its denominator is normalized only over $\mathcal{X}_{\text{neg}}$ rather than the full item universe $\mathcal{X}$, this objective is a sampled and biased approximation to the full-catalog softmax distribution~\citep{blanc2018adaptive}.
At inference time, U2I degenerates to a delta distribution:
\begin{equation}
p_\theta^{\text{u2i-test}}(x \mid c)
=
\begin{cases}
1, & x^{(e)}=e_\theta(c), \\
0, & \text{otherwise}.
\end{cases}
\end{equation}
Accordingly, the recall trigger has only one choice:
\begin{equation}
\tilde{x}^{(e)}=e_\theta(c)
\end{equation}
The recommendation items are then retrieved from $\cal{X}$ via ANN. As seen, U2I concentrates all probability mass on a single retrieval point in the item embedding space. This formulation is overly restrictive, as a user's true interest distribution can be diverse and multi-mode, which lies beyond the representational capacity of a single-point U2I model~\citep{cen2020controllable}.

To address this limitation, prior works such as ComiRec~\citep{cen2020controllable} encode the user context $c$ into multiple triggers during training. The model then selects one of these triggers for contrastive learning according to a matching strategy\footnote{For example, selecting the trigger that is closest to the ground-truth item embedding.}.  At inference time, the model can output multiple triggers, each of which is used to retrieve candidates from the item embedding space. However, this solution still remains within the same U2I framework. It increases only the number of deterministic retrieval points, rather than explicitly modeling the full manifold of the conditional distribution $p(x|c)$. Moreover, the number of user embeddings must be predefined as a hyperparameter, making it difficult to adaptively capture the intrinsic complexity and varying number of modes in a user's interest distribution. Therefore, U2I approaches exhibit poor expressiveness.

Since U2I is only made up of an encoder, its inference cost comes from the feed-forward and self-attention computations over the condition tokens:
\begin{equation}
    \mathcal{C}_{\mathrm{U2I}}^{\text{time}}
    =
    \underbrace{\mathcal{O}(Ld^2N)}_{\text{FFN of condition tokens}}
    +
    \underbrace{\mathcal{O}(LdN^2)}_{\text{attention of condition tokens}}
    \sim
    \mathcal{O}(N^2)
\end{equation}
Moreover, U2I is KV cache-free, since it does not need to maintain KV caches for subsequent computations. Thus, its space complexity is:
\begin{equation}
    \mathcal{C}_{\mathrm{U2I}}^{\text{space}}=\mathcal{O}(1)
\end{equation}

\subsection{SID-AR}
SID-AR~\citep{rajput2023recommender,deng2025onerec,zhou2025onerec} represents each item $x$ as a sequence of SIDs $x^{(s)}_{1:M}$, thereby factorizing the recommendation distribution into an autoregressive sequence model:
\begin{equation}
    p^{\text{SID-AR}}_\theta(x | c)
    =
    \prod_{m=1}^{M}
    p_\theta\left(x^{(s)}_m \mid c,x^{(s)}_{1:m-1}\right)
\end{equation}
This formulation allows the model to be trained in the same way as language models, i.e., using cross-entropy supervision over the SID sequence. At inference time, the retrieval triggers are obtained via beam search:
\begin{equation}
    \tilde{x}_{1:K}^{(s)}=\text{Beamsearch}_K\left(p^{\text{SID-AR}}_\theta(x | c)\right)
\end{equation}
In contrast to U2I, SID-AR defines a compositional distribution over discrete SID sequences and is therefore more flexible in representing multi-mode user interests~\citep{uria2016nade}.

Despite this advantage, the SID-AR formulation introduces two important drawbacks. Firstly, its retrieval quality is highly dependent on the construction of SIDs. Since SIDs are obtained by quantizing continuous item embeddings, the discretization process inevitably discards fine-grained information in the original representation space~\citep{lee2022autoregressive,rajput2023recommender}. Moreover, the discrete code space must be organized such that generated SID sequences can be reliably mapped back to valid and distinguishable items, a requirement that is non-trivial to satisfy at industrial scale. Secondly, SID-AR inherits the efficiency bottleneck of autoregressive decoding. Generating a retrieval trigger requires multiple sequential prediction steps and KV cache maintenance, making the inference cost grow with the SID length and limiting throughput in large-scale retrieval systems~\citep{stern2018blockwise,kwon2023pagedattention}. Specifically, the complexity of SID-AR is:
\begin{equation}
\begin{aligned}
\mathcal{C}_{\mathrm{SID\text{-}AR}}^{\text{time}}&=\underbrace{\mathcal{O}\left(Ld^2N\right)}_{\text{FFN of condition tokens}} + \underbrace{\mathcal{O}\left(LdN^2\right)}_{\text{Attention of condition tokens}}\\
&+\underbrace{\mathcal{O}\left(Ld^2K(M-1)\right)}_{\text{FFN of generated tokens}}+\quad\underbrace{\mathcal{O}\left(\frac{LdK(2N+M-2)(M-1)}{2}\right)}_{\text{Attention of generated tokens}}\sim\mathcal{O}\left(N(N+KM)\right)
\\
\mathcal{C}_{\mathrm{SID\text{-}AR}}^{\text{space}}&=\mathcal{O}\left(L(N+M)\right)
\end{aligned}
\end{equation}
As we can see, compared to U2I, SID-AR has additional time complexity from generated tokens and additional space complexity from maintaining the KV cache, which significantly limits its inference throughput.

Overall, SID-AR approaches suffers from flawed SIDs and low throughput.

\subsection{X-Rec}\label{sec:xrec-formulation}

Combining the advantages of U2I and SID-AR, our proposed X-Rec represents each item $x$ as a continuous embedding $x^{(e)}$ and directly models the recommendation distribution in the continuous embedding space with FM-based diffusion~\citep{lipman2022flow,liu2022flow}. In essence, FM learns a time-dependent velocity field $v_\theta^t(\cdot \mid c)$ that transports a noise sample $\epsilon \sim r(\epsilon)$ toward the target item embedding $x^{(e)}$. During training, the model is optimized with the FM objective, which regresses the predicted velocity field to the ground-truth conditional velocity. During inference, the model generates triggers by denoising a noise step by step, which corresponds to sampling from the learned recommendation distribution\footnote{The form of recommendation distribution $p_\theta^\text{X-Rec}(x|c)$ is implicitly defined by the velocity field: $p_\theta^\text{X-Rec}(x|c)=r(\epsilon)\exp\left(-\int_0^1\nabla\cdot v^t_\theta(\phi^t_\theta(\epsilon|c)|c)dt\right)$, where $\frac{d\phi^t_\theta(\cdot|c)}{dt}=v^t_\theta(\phi^t_\theta(\cdot|c)|c)$~\citep{lai2025principles}.}:
\begin{equation}
    \tilde{x}_{1:K}^{(e)}\sim p_\theta^\text{X-Rec}(x|c)
\end{equation}

\begin{figure}[t]
    \centering
    \includegraphics[width=0.95\textwidth,height=0.47\textheight,keepaspectratio]{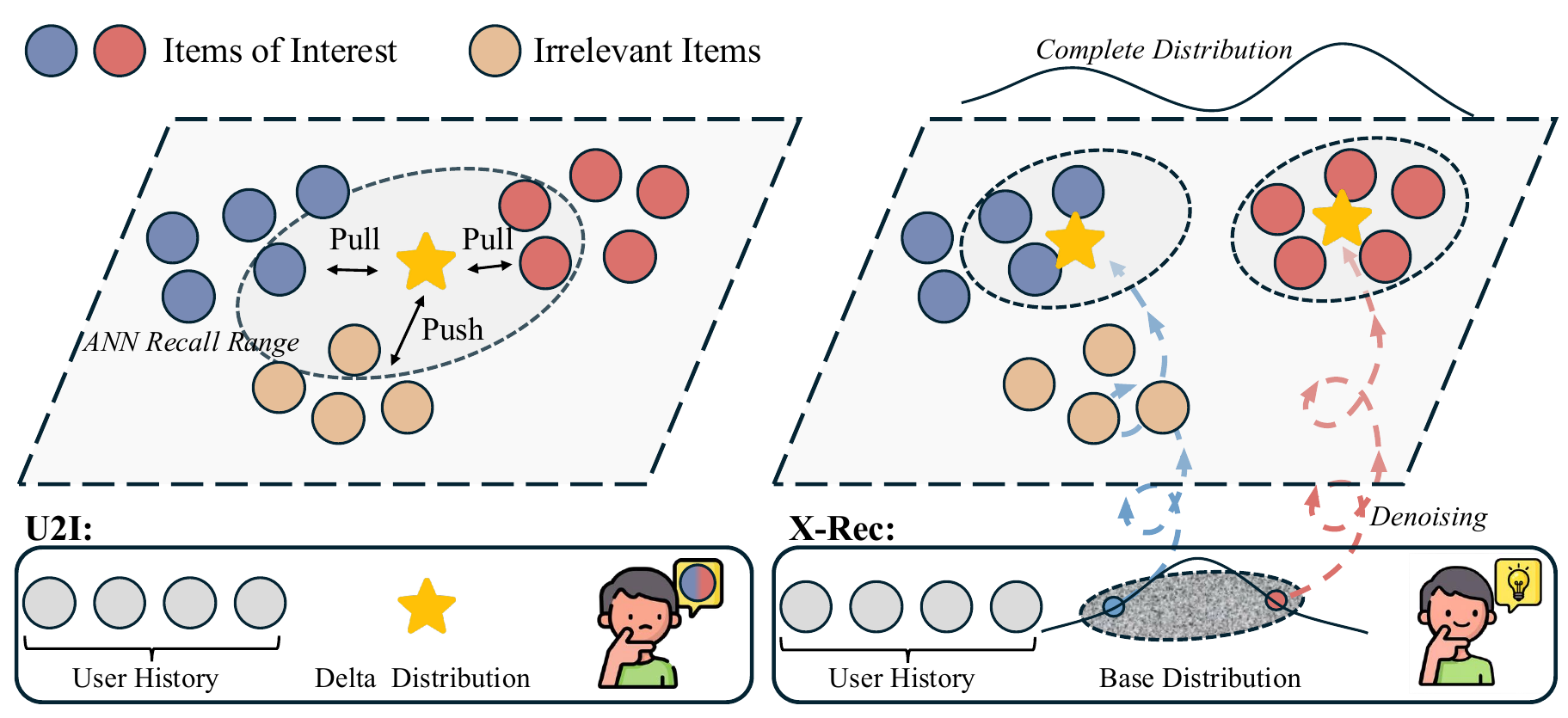}
    \caption{Illustration of the expressiveness of U2I and X-Rec. In the left panel, U2I produces a single retrieval trigger. Its training objective pulls this trigger toward items of interest while pushing it away from irrelevant items. Due to geometric constraints in the embedding space, however, a single trigger can easily fall into a compromised position, making it difficult to retrieve all relevant items without also retrieving irrelevant ones. In the right panel, X-Rec directly models the distribution of items of interest and generates multiple retrieval triggers. These triggers can cover different interest regions, enabling more expressive and precise retrieval of relevant items.}
    \label{fig:u2i-vs-xrec-intuitive}
\end{figure}

Compared with U2I, this formulation can model general continuous distributions, providing sufficient expressiveness to capture diverse and multi-mode user interests, as illustrated in \cref{fig:u2i-vs-xrec-intuitive}\footnote{From a decision-boundary perspective, retrieval with multiple triggers corresponds to selecting candidates \(x \in \mathcal{X}\) such that \(
f(x)=\mathbf{1}\left(\max_{k\in[K]} \left(\tilde{x}_k^{(e)\top} x^{(e)}\right) > \tau \right)\). Thus, the positive retrieval region is a union of \(K\) halfspaces induced by the generated triggers. The VC dimension of this class is \(\Theta(dK\log K)\)~\citep{csikos2019tight}, indicating that increasing the number of triggers directly increases the complexity of the representable decision boundary. This offers a theoretical explanation for the stronger expressiveness of X-Rec over single-vector U2I retrieval: multiple triggers allow X-Rec to represent a richer, multi-region candidate space rather than a single local neighborhood.}. For a case study, please refer to \cref{sec:offline-case-study}.

Compared with SID-AR, X-Rec operates directly on item embeddings, avoiding the quantization errors introduced by discretization and eliminating the need to enforce a fragile one-to-one correspondence between SIDs and item IDs. Meanwhile, the time complexity of X-Rec is:
\begin{equation}
\begin{aligned}
\mathcal{C}_{\text{X-Rec}}^{\text{time}}
=
\underbrace{\mathcal{O}\left(Ld^2N\right)}_{\text{FFN of condition tokens}}
+
\underbrace{\mathcal{O}\left(LdN^2\right)}_{\text{attention of condition tokens}} +
\underbrace{\mathcal{O}\left(L'Td^2K\right)}_{\text{FFN of generated embeddings}}
+
\underbrace{\mathcal{O}\left(L'TdKN\right)}_{\text{attention of generated embeddings}}
\end{aligned}
\end{equation}
Here, \(L'\) denotes the number of Transformer layers used in each velocity field evaluation, and \(T\) denotes the number of denoising steps. As shown in \cref{sec:effect-late-interaction}, we empirically find that using only the final Transformer layer at each denoising step, i.e., \(L'=1\), and setting the number of denoising steps equal to the total number of Transformer layers, i.e., \(T=L\), provides a favorable quality--efficiency trade-off. Under this efficient configuration, the time complexity of X-Rec becomes:
\begin{equation}
\begin{aligned}
\mathcal{C}_{\text{X-Rec}}^{\text{time}}=
\underbrace{\mathcal{O}\left(Ld^2N\right)}_{\text{FFN of condition tokens}}
+
\underbrace{\mathcal{O}\left(LdN^2\right)}_{\text{attention of condition tokens}}+
\underbrace{\mathcal{O}\left(Ld^2K\right)}_{\text{FFN of generated embeddings}}
+
\underbrace{\mathcal{O}\left(LdKN\right)}_{\text{attention of generated embeddings}}\\
\sim
\mathcal{O}\left(N(N+K)\right)
\end{aligned}
\end{equation}
With this design, X-Rec only need to store the last-layer KV cache of the condition tokens, i.e.,
\begin{equation}
    \mathcal{C}_{\text{X-Rec}}^{\text{space}}=\mathcal{O}(N)
\end{equation}
Therefore, the complexity of X-Rec lies between U2I and SID-AR. More importantly, X-Rec achieves a favorable trade-off between expressiveness and efficiency: it significantly improves upon U2I by modeling a flexible continuous recommendation distribution, while avoiding the expensive token-by-token autoregressive decoding required by SID-AR. This makes X-Rec expressive enough to capture complex user interest distributions and efficient enough for large-scale retrieval.

\section{X-Rec Framework}\label{sec:x-rec_framework}
In this section, we present the proposed X-Rec framework in detail. We begin with a brief overview of FM, then describe the high-level training and inference algorithms of X-Rec, followed by the concrete implementation of the model architecture. Finally, we introduce the overall training pipeline.

\subsection{Preliminary: Flow Matching}\label{sec:flow-matching}

Given a recommendation context \(\hat{c}\), let \(\hat{x}^{(e)}\in\mathcal{X}^{(e)}\) denote the item embedding of the ground truth item \(\hat{x}\)\footnote{In this paper, $\hat{\cdot}$ indicates a sample drawn from an empirical distribution.}, sampled from an empirical recommendation distribution \(p_{\mathrm{data}}(x \mid \hat{c})\). The key idea of FM is to learn a time-dependent flow function \(\phi_\theta^t(\cdot \mid \hat{c})\), where \(t\in[0,1]\), that transforms a simple continuous random variable \(\epsilon\sim r(\epsilon)\), i.e., noise, into a sample from the target recommendation distribution~\citep{lipman2023flow}. Specifically, the transformed random variable \(\phi_\theta^1(\epsilon \mid \hat{c})\) induces a model distribution \(p_\theta(x^{(e)} \mid \hat{c})\), which is expected to approximate \(p_{\mathrm{data}}(x^{(e)} \mid \hat{c})\). Therefore, the learned flow should satisfy
\begin{equation}
    \phi_\theta^0(\epsilon \mid \hat{c})=\epsilon\sim r(\epsilon),
    \quad
    \phi_\theta^1(\epsilon \mid \hat{c})\sim p_{\mathrm{data}}(x^{(e)} \mid \hat{c})
    \label{eq:flow_requirements}
\end{equation}
Once such a flow is learned, we can sample from \(p_\theta\left(x^{(e)} \mid \hat{c}\right)\) by first drawing noise from \(r(\epsilon)\) and then denoising it through the learned flow. However, directly constructing a flow that satisfies the above requirement is challenging. Fortunately, it is straightforward to define a conditional flow when the target embedding \(\hat{x}^{(e)}\) is given. This conditional flow \(\phi^t(\epsilon \mid \hat{x}^{(e)},\hat{c})\) can be constructed by interpolating from the noise \(\epsilon\) to the target embedding \(\hat{x}^{(e)}\) over time \(t\):
\begin{equation}
    \phi^0(\epsilon \mid \hat{x}^{(e)},\hat{c})=\epsilon\sim r(\epsilon),
    \quad
    \phi^1(\epsilon \mid \hat{x}^{(e)},\hat{c})=\hat{x}^{(e)}
\end{equation}
In most cases, the conditional flow is independent of \(\hat{c}\). For example, Euclidean FM can use the linear interpolation \(\phi^t(\epsilon \mid \hat{x}^{(e)},\hat{c})=(1-t)\epsilon+t\hat{x}^{(e)}\)~\citep{lipman2023flow}. For convenience, we abbreviate \(\phi^t(\epsilon \mid \hat{x}^{(e)},\hat{c})\) as \(\bar{x}^{(e)}_t\). Taking the derivative of the conditional flow with respect to time yields the corresponding conditional velocity field:
\begin{equation}
    v^t\left(\bar{x}^{(e)}_t \mid \hat{x}^{(e)},\hat{c}\right)
    =
    \frac{d}{dt}\bar{x}^{(e)}_t
\end{equation}
The FM objective then regresses a neural velocity field \(v_\theta^t(\bar{x}_t^{(e)}|\hat{c})\) to match this conditional velocity field in expectation over data, noise, and timestep:
\begin{equation}
\begin{aligned}
    \mathcal{L}_{\mathrm{FM}}
    =
    \mathbb{E}_{\substack{
        \hat{c},\hat{x}^{(e)}\sim p_{\mathrm{data}},\\
        \epsilon\sim r,\,
        t\sim \mathcal{U}(0,1)
    }}
    \left[
    \left\Vert
    v_\theta^t\left(
        \bar{x}^{(e)}_t
        \mid \hat{c}
    \right)
    -
    \frac{d}{dt}\bar{x}^{(e)}_t
    \right\Vert_{\mathcal{X}^{(e)}}^2
    \right]
\end{aligned}\label{eq:fm_loss}
\end{equation}
\cite{lipman2023flow} have proven that the optimal regressed velocity field \(v_{\theta^\star}^t(\cdot \mid \hat{c})\) induces a flow function that satisfies the construction requirement in \cref{eq:flow_requirements}. More precisely, the induced flow \(\phi_{\theta^\star}^t(\epsilon \mid \hat{c})\), which we refer to as \(\tilde{x}_t^{(e)}\), is defined as the solution to the following ordinary differential equation (ODE)\footnote{In this paper, $\tilde{\cdot}$ indicates a model-estimated counterpart of the underlying random variable.}:
\begin{equation}
\begin{aligned}
    \frac{d}{dt}
    \tilde{x}_t^{(e)}
    &=
    v_{\theta^\star}^t
    \left(
        \tilde{x}_t^{(e)}
        \mid \hat{c}
    \right)\\
    \tilde{x}_0^{(e)}
    &= \epsilon \sim r(\epsilon)
\end{aligned}
\end{equation}
At inference time, we numerically integrate the above learned flow on the representation space \(\mathcal{X}^{(e)}\) using Euler discretization, as we shall see in \cref{sec:inference}.

\subsection{Algorithm}\label{sec:algorithm}
This section presents the learning and generation procedures of X-Rec. During training, the model jointly learns coarse semantic anchor prediction and anchor-conditioned FM on the hypersphere. During inference, X-Rec first identifies an anchor from the recommendation context and then integrates the learned vector field on the hypersphere to produce recall triggers. We describe the training and inference stages in turn.

\subsubsection{Training}

\paragraph{Anchor Loss.}
Although FM provides a general simulation-free framework for learning continuous transport paths~\citep{lipman2023flow}, directly generating fine-grained item embeddings remains challenging due to the complexity and multi-mode nature of the item representation space. To address this issue, we introduce anchor conditioning, which enables a coarse-to-fine generation process. Concretely, X-Rec first predicts a coarse semantic region of the target item and then generates the recall-trigger embedding conditioned on this region. We represent the coarse region by the cluster-center index \(\hat{x}^{(a)}\in\mathcal{A}\) of the target item embedding \(\hat{x}^{(e)}\). We refer to \(\hat{x}^{(a)}\) as an ``anchor'' because it provides a coarse localization of the target embedding. This decomposition reduces the difficulty of continuous embedding generation and thus makes the generation process more stable and controllable.

Naturally, we optimize anchor prediction with the cross-entropy loss:
\begin{equation}
    \mathcal{L}_{\mathrm{anchor}}
    =
    -
    \mathbb{E}_{(\hat{c},\hat{x}^{(a)})\sim p_{\mathrm{data}}}
    \left[
        \log p_\theta(\hat{x}^{(a)}\mid \hat{c})
    \right]
\end{equation}
where \(\hat{x}^{(a)}\) denotes the ground-truth anchor of the target item under context \(\hat{c}\).

\paragraph{RFM Loss.} Conditioned on both the user context and anchor, X-Rec then generates the final recall trigger by progressively denoising the noise. As aforementioned, each item \(\hat{x}\in\mathcal{X}\) is represented by a embedding \(\hat{x}^{(e)}\in\mathcal{X}^{(e)}\). The embedding model is trained with contrastive representation learning and \(\ell_2\) normalization, so item embeddings naturally lie on the unit hypersphere, i.e., \(\mathcal{X}^{(e)}=\mathcal{S}^{d_e-1}\), where \(d_e\) is the embedding dimension. In this setting, directly applying FM in Euclidean space using linear interpolation and additive denoising may cause intermediate and generated states deviate from the hypersphere. Such off-manifold generated triggers are misaligned with the normalized item embeddings. Therefore, we instantiate the general formulation in Section~\ref{sec:flow-matching} as Riemannian Flow Matching (RFM) (\cite{chen2024flowgeometry,kumar2026learning}), ensuring that the flow trajectory remains on the hypersphere.

Specifically, RFM constructs the conditional flow through geodesic interpolation:
\begin{equation}
    \bar{x}_t^{(e)}
    =
    \frac{\sin((1-t)\omega)}{\sin\omega}\epsilon
    +
    \frac{\sin(t\omega)}{\sin\omega}\hat{x}^{(e)},
    \quad
    \omega=\arccos(\epsilon^\top\hat{x}^{(e)})
\end{equation}
where \(\epsilon\sim r(\epsilon)\) is sampled from the base distribution on the hypersphere, and \(\omega\) is the geodesic distance between \(\epsilon\) and \(\hat{x}^{(e)}\). The corresponding conditional velocity field is obtained by differentiating the geodesic path with respect to \(t\):
\begin{equation}
    \frac{d}{dt}
    \bar{x}_t^{(e)}
    =
    \frac{\omega}{\sin\omega}
    \left(
        \cos(t\omega)\hat{x}^{(e)}
        -
        \cos((1-t)\omega)\epsilon
    \right)
\end{equation}
Substituting this conditional velocity into the flow-matching objective in Equation (\ref{eq:fm_loss}), we can then obtain the final RFM loss.
\begin{equation}
\begin{aligned}
    \mathcal{L}_{\mathrm{RFM}}
    =
    \mathbb{E}_{\substack{
        (\hat{c},\hat{x}^{(a)},\hat{x}^{(e)})\sim p_{\mathrm{data}},\\
        \epsilon\sim r(\epsilon),\,\\
        t\sim \mathcal{U}(0,1)
    }}
    \Bigg[
    \Bigg\|
    v_\theta^t
    \left(
        \frac{\sin((1-t)\omega)}{\sin\omega}\epsilon
        +
        \frac{\sin(t\omega)}{\sin\omega}\hat{x}^{(e)}
        \mid \hat{c},\hat{x}^{(a)}
    \right)
    -
    \frac{\omega}{\sin\omega}
    \left(
        \cos(t\omega)\hat{x}^{(e)}
        -
        \cos((1-t)\omega)\epsilon
    \right)
    \Bigg\|^2
    \Bigg]
    \\
    \quad
    \omega=\arccos(\epsilon^\top\hat{x}^{(e)})
\end{aligned}
\end{equation}

\paragraph{TotalLoss.} In practice, for each training example, we sample multiple noise vectors $\epsilon$ and timesteps $t$, and use the resulting empirical mean to estimate the expectation in the RFM objective. The final training objective of X-Rec is a weighted sum of the above two terms:
\begin{equation}
    \mathcal{L}=\alpha\mathcal{L}_{\text{anchor}}+\beta\mathcal{L}_{\text{RFM}}
    \label{eq:xrec_loss}
\end{equation}
where $\alpha$ and $\beta$ are weighting coefficients.

\subsubsection{Inference}\label{sec:inference}
During inference, X-Rec first predicts an anchor $\tilde{x}^{(a)} \sim p_\theta(x^{(a)}\mid \hat{c})$. Then, it initializes the state from the base distribution $\tilde{x}^{(e)}_0\sim r(\epsilon)=U_{\mathcal{S}^{d_e-1}}(\epsilon)$, where \(U_{\mathcal{S}^{d_e-1}}(\epsilon)\) denotes the uniform distribution on the unit hypersphere. Next, it denoises this state progressively via Riemannian Euler integration with timestep interval $\Delta t$:
\begin{equation}
\tilde{x}^{(e)}_{t+\Delta t}
=
\cos\left(
    \left\|
    v_\theta^t(\tilde{x}^{(e)}_t\mid \hat{c},\tilde{x}^{(a)})
    \right\|
    \Delta t
\right)
\tilde{x}^{(e)}_t +
\sin\left(
    \left\|
    v_\theta^t(\tilde{x}^{(e)}_t\mid \hat{c},\tilde{x}^{(a)})
    \right\|
    \Delta t
\right)
\frac{
    v_\theta^t(\tilde{x}^{(e)}_t\mid \hat{c},\tilde{x}^{(a)})
}{
    \left\|
    v_\theta^t(\tilde{x}^{(e)}_t\mid \hat{c},\tilde{x}^{(a)})
    \right\|
}
\end{equation}
Once \(t\) reaches \(1\), the terminal state \(\tilde{x}^{(e)}_1\) is used as the final recall trigger.

\subsection{Model Architecture}
\begin{figure}[t]
    \centering
    \includegraphics[width=0.9\textwidth,height=1.0\textheight,keepaspectratio]{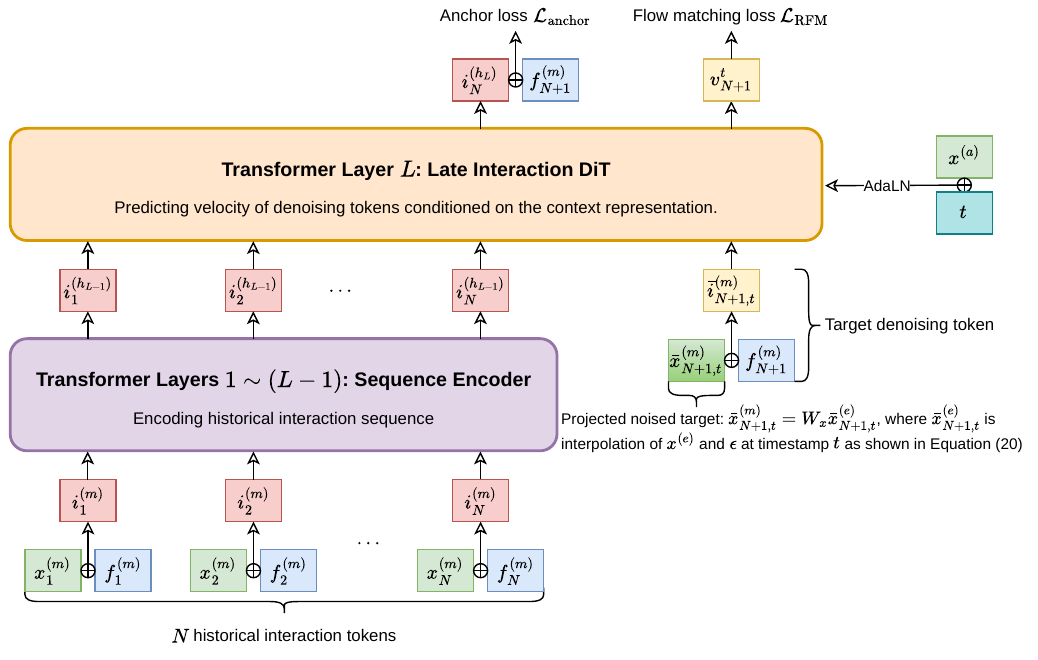}
    \caption{Model architecture of X-Rec. Historical interaction tokens are first encoded by the first \(L-1\) Transformer layers, and the target denoising tokens are then fed only into the last Transformer layer to predict the velocity. Pretraining and SFT share the same architecture but differ in context--target construction and attention visibility.}
    \label{fig:xrec-architecture}
\end{figure}

As introduced in ~\cref{sec:algorithm}, X-Rec is consists of two key modules: an anchor predictor \(p_\theta(x^{(a)} \mid c)\) and a velocity predictor \(v_\theta^t(\tilde{x}_t^{(e)} \mid c, x^{(a)})\). We first describe how the features in the recommendation context are encoded, and then detail the architectures of the two modules.

\subsubsection{Feature Encoding}
The recommendation context \(c\) consists of the historical interaction sequence \(s\) and the expected target attribute \(f\). We now describe how \(s\) and \(f\) are represented in our framework. First, the attribute is represented by projecting its sub-attribute embeddings to the model dimension $d_m$:
\begin{equation}
    f^{(m)}=W_ff^{(e)}=W_f[{f^1}^{(e)};
        {f^2}^{(e)};
        \cdots;
        {f^F}^{(e)}]\in\mathbb{R}^{d_m}
\end{equation}
where $W_f$ is the leanable attribute projection matrix. Each interaction in the historical interaction sequence $s$, i.e., a tuple \(i=(x,f)\in h\), is represented by fusing the information of item and attributes:
\begin{equation}
    i^{(m)}=x^{(m)}+f^{(m)}=W_xx^{(e)}+W_ff^{(e)}\in\mathbb{R}^{d_m}
\end{equation}
where $W_x$ is the item-embedding projection matrix. The encoded historical interaction sequence $s$ is therefore represented as a sequence of interaction tokens:
\begin{equation}
    s^{(m)}
    =
    \left(
        i^{(m)}_1,
        i^{(m)}_2,
        \cdots,
        i^{(m)}_N
    \right)
    \in \mathbb{R}^{d_m\times N}
\end{equation}
As illustrated in \cref{fig:xrec-architecture}, the encoded interaction sequence \(s^{(m)}\) is then fed into an \(L\)-layer Transformer to produce multi-level contextualized user representations:
\begin{equation}
\begin{aligned}
    s^{(h_1)}&=\operatorname{TransformerLayer}_1(s^{(m)})=(i_1^{(h_1)},i_2^{(h_1)},\cdots,i_N^{(h_1)})\in\mathbb{R}^{d_m\times N}\\
    s^{(h_2)}&=\operatorname{TransformerLayer}_2(s^{(h_1)})=(i_1^{(h_2)},i_2^{(h_2)},\cdots,i_N^{(h_2)})\in\mathbb{R}^{d_m\times N}\\
    \cdots \\
    s^{(h_L)}&=\operatorname{TransformerLayer}_L(s^{(h_{L-1})})=(i_1^{(h_L)},i_2^{(h_L)},\cdots,i_N^{(h_L)})\in \mathbb{R}^{d_m\times N}
\end{aligned}
\end{equation}
We adopt the Qwen3-style (\cite{yang2025qwen3technicalreport,qwen35blog}) Transformer backbone, where RoPE (\cite{su2024roformer}) is used for positional encoding and QK-normalization is applied to stabilize training. The user sequence is encoded with a causal attention mask, making \(s^{(h_L)}_N\) a valid contextualized representation of the historical interaction sequence. Moreover, this masking strategy allows the resulting KV cache to be reused during the subsequent velocity-prediction stage, which significantly improves inference throughput.

\subsubsection{Anchor Predictor}
Since the anchor $x^{(a)}$ is discrete, the anchor predictor is naturally a softmax classifier based on the historical-sequence embedding \(i^{(h_L)}_N\) and the expected target-attribute embedding \(f^{(m)}\):
\begin{equation}
    p_\theta(x^{(a)}|c)=\frac{\exp\left(w_a^\top(i^{(h_L)}_N+f^{(m)})\right)}{\sum_{a'\in\mathcal{A}}\exp\left(w_{a'}^\top(i^{(h_L)}_N+f^{(m)})\right)}, \qquad w_a\in\mathbb{R}^{d_m}
\end{equation}
where $w_a$ is the classifier weight vector for anchor $a$.

\subsubsection{Velocity Predictor: Late Interaction DiT}
The velocity predictor \(v_\theta^t(\tilde{x}_t^{(e)}\mid c,x^{(a)})\) largely follows the DiT design~\citep{peebles2023scalable,wu2025qwen}, with one important modification: we introduce a late-interaction~\citep{khattab2020colbert} structure for efficient velocity computation. Given the current denoising state \(\tilde{x}^{(e)}_t\), we first incorporate the target attribute information \(f^{(m)}\) into the denoising state in the same way of the interaction tokens:
\begin{equation}
    \tilde{i}_t^{(m)}=
    \tilde{x}_t^{(m)}+f^{(m)}=W_x\tilde{x}^{(e)}_t+W_ff^{(e)}\in \mathbb{R}^{d_m}
\end{equation}
Instead of feeding the denoising token \(\tilde{i}_t^{(m)}\) through all Transformer layers, we insert it only into the last layer. This design allows the model to reuse the contextualized historical representations computed in the prefill stage, while using the final layer to estimate the velocity field for the current denoising state. Formally,
\begin{equation}
\begin{aligned}
    \left[
        s^{(h_L)};
        \tilde{i}^{(h_L)}_t
    \right]
    &=
    \operatorname{TransformerLayer}_L
    \left(
        \left[
            s^{(h_{L-1})};
            \tilde{i}_t^{(m)}
        \right],
        t,
        x^{(a)}
    \right)
    \in \mathbb{R}^{(N+1)\times d_m}\\
    {v'}_\theta^t
    \left(
        \tilde{x}_t^{(e)}
        \mid c,x^{(a)}
    \right)
    &=
    W_v \tilde{i}^{(h_L)}_t
    \in \mathbb{R}^{d_e}\\
    v_\theta^t
    \left(
        \tilde{x}_t^{(e)}
        \mid c,x^{(a)}
    \right)
    &=
    \underbrace{
    {v'}_\theta^t
    \left(
        \tilde{x}_t^{(e)}
        \mid c,x^{(a)}
    \right)
    -
    \left(
        {v'}_\theta^t
        \left(
            \tilde{x}_t^{(e)}
            \mid c,x^{(a)}
        \right)^\top
        \tilde{x}_t^{(e)}
    \right)
    \tilde{x}_t^{(e)}
    }_{\text{projection onto }T_{\tilde{x}_t^{(e)}}\mathcal{S}^{d_e-1}}
    \in
    T_{\tilde{x}_t^{(e)}}\mathcal{S}^{d_e-1}
\end{aligned}
\end{equation}
where \(\big(\left({v'}_\theta^t\right)^\top\tilde{x}_t^{(e)}\big)\tilde{x}_t^{(e)}\) is the radial velocity component along the current state, and \(T_{\tilde{x}_t^{(e)}}\mathcal{S}^{d_e-1}=\{u\in\mathbb{R}^{d_e}:u^\top\tilde{x}_t^{(e)}=0\}\) denotes the tangent space of the unit hypersphere at \(\tilde{x}_t^{(e)}\). The projection removes the radial component of the velocity, ensuring that the predicted velocity lies in the tangent space.

As illustrated in \cref{fig:xrec-architecture}, the flow timestep \(t\) and anchor \(x^{(a)}\) condition the final Transformer layer through AdaLN~\citep{peebles2023scalable}: their embeddings are added up, and jointly modulate the attention and MLP blocks. A causal mask is applied when encoding the historical interaction sequence, allowing its KV cache to be reused across iterative velocity evaluations. As shown in \cref{sec:effect-late-interaction}, this late-interaction design substantially improves generation throughput, with only a marginal decrease in \(\operatorname{Recall}@20\) compared with full-layer denoising.

\subsection{Training Pipline}\label{sec:training-inference-protocols}
Training X-Rec includes two stages: pretraining and supervised fine-tuning (SFT). The pretraining stage aims to capture general sequential transition structure and knowledge from users' historical interaction sequences and item embeddings, enabling the model to learn broadly transferable representations. The SFT stage further adapts the pretrained model to specific recommendation scenarios, such as particular content types or expected target attributes.

\begin{figure}[t]
    \centering
    \includegraphics[width=1.0\textwidth,height=1.0\textheight,keepaspectratio]{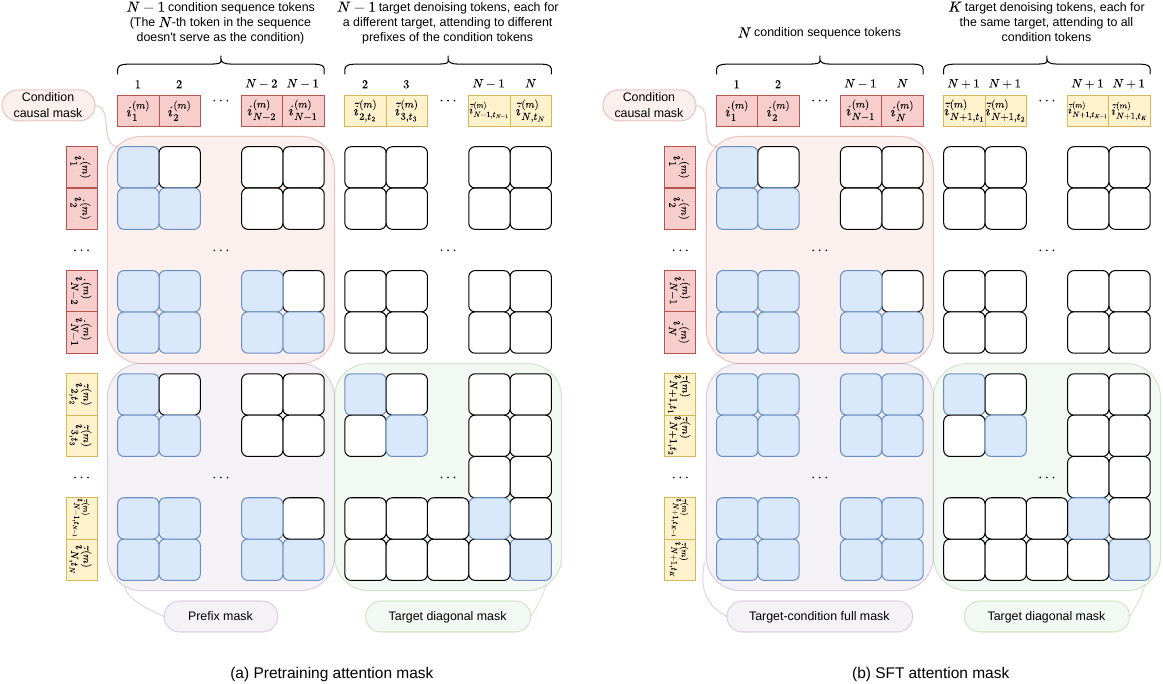}
    \caption{Attention masks used in X-Rec. During pretraining, each denoising token corresponds to a different target and attends only to its designated history prefix and itself. Accordingly, the position index of each denoising token is set to the length of its corresponding prefix plus one. During SFT, all denoising tokens correspond to the same target and attend to the complete condition sequence. Their position indices are therefore identical, i.e., the length of the condition sequence plus one. In both stages, condition tokens are encoded with a causal mask, allowing the last-layer KV cache to be reused across denoising tokens at different timesteps. Meanwhile, denoising tokens remain mutually invisible to one another.}
    \label{fig:xrec-attention-masks}
\end{figure}

\subsubsection{Pretraining}\label{sec:pretrain}
Given a chronological user historical interaction sequence \(s=(i_1,i_2,\cdots,i_N),i_n=(x_n,f_n)\), the pretraining stage models the following autoregressive factorization of the item distribution:
\begin{equation}
    \prod_{n=1}^Np_\theta(x_n|i_{1:n-1},f_n)
\end{equation}
Here, each factor $p_\theta(x_n|i_{1:n-1},f_n)$ can be optimized using the X-Rec loss. Thus, a single sequence yields $N$ next-item prediction pairs according to different prefix lengths.

Specifically, for each target item \(x_n\), the model samples a noise vector and a timestep, and evaluates the loss defined in \cref{eq:xrec_loss}. To improve training efficiency, we introduce a specialized attention mask that allows the losses for all sequence positions to be computed within a single prefill pass. As illustrated in \cref{fig:xrec-attention-masks}, this mask restricts each denoising token to attend only to its corresponding historical prefix, while keeping different denoising tokens isolated from one another. As a result, X-Rec can train all prefix--target pairs in parallel without changing the semantics of sequential pretraining.

After pretraining, the model captures general item-transition patterns. In the next stage, SFT, it further adapts the context and target semantics to the downstream serving task.
\subsubsection{Supervised Fine-Tuning}\label{sec:sft}

SFT adapts the pretrained transition model to the retrieval events used by the downstream service. For each eligible target item \(x\) with attribute \(f\), we construct its sequence \(s\) by chronologically concatenating two types of user history: a historical positive sequence and a recent interaction sequence. The historical positive sequence consists of items with which the user has had meaningful engagement, whereas the recent interaction sequence contains the user's most recent activity before the target event. These two sources provide complementary signals: the former reflects stable and persistent user preferences, while the latter captures immediate and transient intent that may not appear in the long-term positive history~\citep{yu2019adaptive}.

For each target item, we draw eight independent noise samples and timesteps to estimate the RFM objective. As shown in Figure \cref{fig:xrec-attention-masks}, we also use a specialized attention mask, where each denoising token attends to the complete composed history and the same expected target attribute, while different noised copies remain mutually isolated. This allows multiple denoising tokens for the same target to be trained in parallel without introducing information leakage across replicas.

Overall, the SFT stage adapts the pretrained model to the downstream retrieval task, enabling it to generate retrieval triggers that better match the serving-time user context and target attribute.

\section{Offline Experiments}\label{sec:offline-experiments}

In this section, we evaluate X-Rec in offline settings. We first describe the experimental setup, including the chronological streaming benchmark, evaluation metrics, and baselines. We then compare X-Rec with representative methods and analyze its key design choices through ablation studies.

\subsection{Experimental Settings}\label{sec:experimental-settings}

\subsubsection{Data}\label{sec:offline-data-protocol}

We evaluate X-Rec on an industrial streaming benchmark constructed from TikTok platform. All models are first pretrained on large-scale historical interaction sequences and then trained and evaluated on the stream, following the continual-update setting used in production. For pretraining and streaming data, the item attribute set \(f\) includes only user actions. Therefore, the task is to generate items that are aligned with the user's demonstrated interests, conditioned on the user's historical interaction sequence.

\paragraph{Pretraining Data.}
We construct pre-training data from large-scale user–item interaction logs on the platform. The sequences are divided into chunks containing at most 200 interactions.

\paragraph{Streaming Evaluation Benchmark.}
To approximate production retrieval, where models are continuously updated with fresh user feedback, we construct a streaming benchmark comprising 47 consecutive partitions. Each partition includes 2.6M sampled training examples and 10K test examples, where each example consists of a historical positive-interaction sequence and a recent-interaction sequence, both of bounded length. We collect the target items from each partition to construct the corresponding ANN recall source. Models are trained and evaluated sequentially over the partitions: the final checkpoint from one partition is used to initialize training on the next, preserving the temporal order of both data and model updates. This protocol captures model performance under continual adaptation and provides a more faithful offline approximation of online retrieval behavior than a static evaluation split.

\subsubsection{Evaluation Metrics}\label{sec:offline-metrics}

For each test example \((c,x)\), the model generates \(K\) embedding triggers \(t_{1:K}\). Each trigger is used to retrieve its top-\(R\) nearest items from the ANN index. We define \(\operatorname{Recall}@\left(K{\times}R\right)\) as
\begin{equation}
    \operatorname{Recall}@\!\left(K{\times}R\right)
    =
    \frac{1}{|\mathcal{D}|}
    \sum_{(c,x)\in \mathcal{D}}
    \mathbb{I}
    \left[
        x
        \in
        \bigcup_{k=1}^{K}
        \operatorname{ANN}_{R}\!\left(t_k\right)
    \right],
\end{equation}
where \(\operatorname{ANN}_{R}(t)\) denotes the set of top-\(R\) items retrieved by trigger \(t\), and \(\mathcal{D}\) denotes the testing set. This metric measures the fraction of test examples for which the ground-truth item is included in the union of retrieved candidates. By default, we report \(\operatorname{Recall}@20{\times}1\), which we abbreviate as \(\operatorname{Recall}@20\). \cref{app:xrec-per-snapshot-recall} reports the complete per-snapshot results.

We also evaluate generation efficiency using throughput, measured as requests per second on a single production-grade GPU. Each request generates \(K=20\) triggers, matching the default retrieval setting. The throughput measurement includes only trigger generation and excludes the ANN lookup stage.

\subsubsection{Baseline Methods}\label{sec:offline-baselines}

We compare X-Rec against a broad range of representative open-source baselines from two categories. The first category comprises U2I methods, including SASRec~\citep{kang2018self}, LRURec~\citep{yue2024lrurec}, and DreamRec~\citep{yang2023dreamrec}\footnote{Although DreamRec is diffusion-based, its original implementation generates only a single embedding for retrieval. We therefore follow this setting and categorize DreamRec as a U2I method.}. The second category comprises SID-based generative methods, including TIGER~\citep{rajput2023recommender}, EAGER~\citep{wang2024eager}, LATTE~\citep{hou2026latte}, and LLaDARec~\citep{shi2025lladarec}. For all open-source baselines, we follow the model configurations reported in their original papers.

We additionally implement two controlled variants of X-Rec to disentangle the effects of the learning paradigm and generation architecture. X-Rec-U2I is trained with a contrastive objective and produces a single trigger embedding for retrieval, whereas X-Rec-AR employs a decoder-only architecture to generate SIDs autoregressively. Both variants use the same context-encoder configuration as X-Rec, as detailed in \cref{app:xrec-variant-designs}.

For U2I methods, which produce a single retrieval embedding, we compute \(\operatorname{Recall}@20\) under the \(1{\times}20\) setting. For SID-based methods and X-Rec, which support multiple retrieval triggers, we use the \(20{\times}1\) setting by default. We denote both settings as \(\operatorname{Recall}@20\) when the underlying configuration is clear from context.

All models follow the same pretraining-then-streaming-SFT pipeline. To ensure a consistent comparison between SID-based and embedding-based retrieval methods, we adopt a unified ANN evaluation protocol. Specifically, the SIDs generated by SID-based methods are decoded into embeddings through their respective RQ codebooks, and the resulting embeddings are used as triggers for ANN retrieval.

\subsubsection{Implementations}\label{sec:offline-implementation}

X-Rec adopts Qwen3-0.6B~\citep{yang2025qwen3technicalreport} as its Transformer backbone. After removing the token embedding layer, the resulting model contains approximately 0.46B parameters. Each item is a video, and is represented by a 128-dimensional multimodal embedding normalized to the unit hypersphere. SIDs are obtained by quantizing item embeddings with K-means residual quantization using \(4\) codebooks of size \(4096\)~\citep{zhou2025onerectechnicalreport}, yielding an average reconstruction cosine similarity of \(0.94\).

During pretraining, we train the model for one epoch (200K steps) using AdamW~\citep{loshchilov2017decoupled} with a batch size of 4096. The learning rate is linearly warmed up to \(5\times10^{-4}\) over the first 2K steps and then decayed to \(1\times10^{-4}\) following a cosine schedule. During streaming SFT, we train the model for three epochs using AdamW with a constant learning rate of \(1\times10^{-4}\) and a batch size of 512. The maximum length of the user interaction history is set to 200 in both stages. We set the loss coefficients to \(\alpha=0.1\) and \(\beta=1\). To enable classifier-free guidance (CFG)~\citep{ho2022classifierfree} through marginal velocity estimation, we randomly drop the conditioning context with probability 5\% during training.

At inference time, X-Rec first samples anchors with a temperature of 1.4 and then generates retrieval triggers over 30 denoising steps. We apply CFG with a guidance scale of 3.0 throughout the denoising process.

\subsection{Overall Performance}\label{sec:offline-overall-performance}

\begin{figure}[t]
    \centering
    \includegraphics[width=\textwidth]{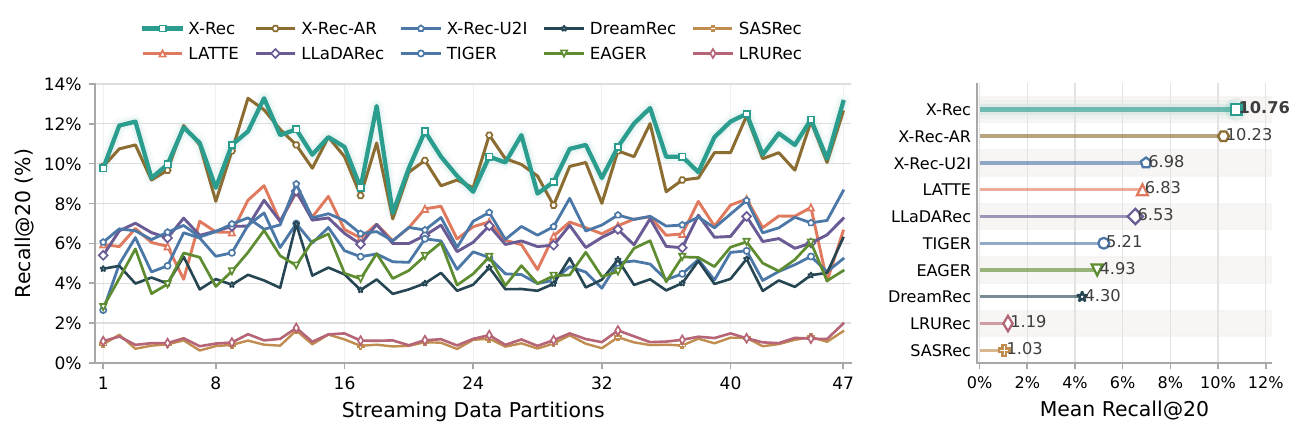}
    \caption{Comparison of different models on the streaming benchmark. The left panel shows the results for each partition, while the right panel ranks the methods by their average performance.}
    \label{fig:streaming-baseline-comparison}
\end{figure}

As shown in \cref{fig:streaming-baseline-comparison}, on our streaming benchmark, X-Rec substantially outperforms most baselines. After controlling for model configuration with the two X-Rec variants, we make the following two main observations.

First, X-Rec achieves slightly higher recall than X-Rec-AR. On average, X-Rec reaches \(10.65\%\), outperforming X-Rec-AR by \(0.53\%\) percentage points. Across the 47 streaming partitions, X-Rec performs better than X-Rec-AR on 38 partitions, with one tie. This suggests that, in terms of distribution modeling capacity, FM and SID-AR are comparable. More importantly, as we will show later, X-Rec achieves this performance with substantially higher generation throughput, since its triggers can be generated in parallel rather than decoded sequentially token-by-token.

Second, X-Rec clearly outperforms X-Rec-U2I. X-Rec-U2I achieves an average recall of only \(6.98\%\), trailing X-Rec by \(3.67\%\) percentage points. Since X-Rec-U2I relies on a single retrieval embedding, the gap highlights the importance of multi-mode interest modeling. X-Rec addresses this by generating multiple triggers for different interest modes. Moreover, as shown in \cref{sec:offline-trigger-depth-scaling}, even when X-Rec generates only one trigger and is evaluated under the same \(\operatorname{Recall}@1{\times}20\) protocol as X-Rec-U2I, it still achieves better performance.

\subsection{Ablation Study}\label{sec:offline-ablation}
We further conduct ablation studies to examine the contributions of several key components, including anchor conditioning, RFM, late interaction, and pretraining.

\subsubsection{Effect of Anchor Conditioning}\label{sec:offline-anchor-ablation}

\begin{figure}[t]
    \centering
    \includegraphics[width=\textwidth]{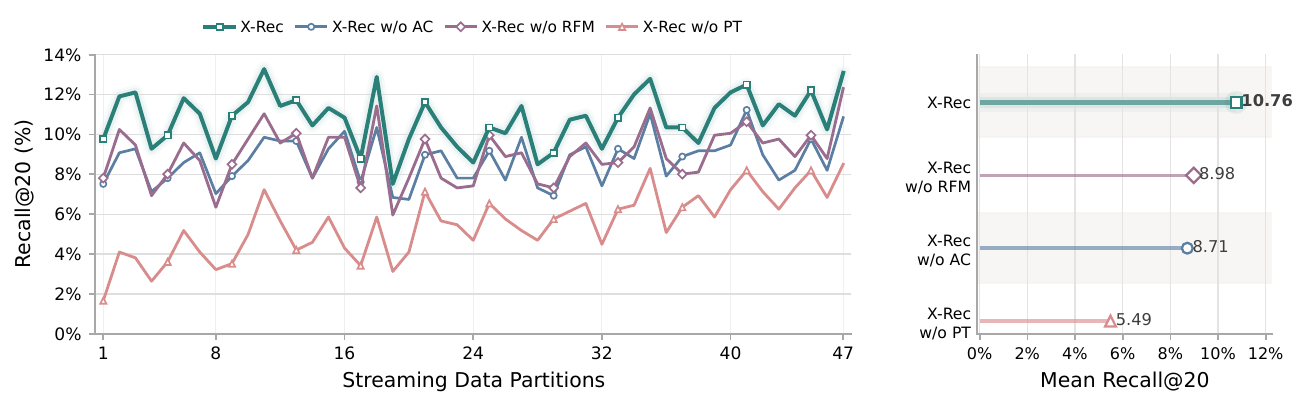}
    \caption{Ablation results of anchor conditioning (AC), RFM, and pretraining (PT) on the streaming benchmark.}
    \label{fig:xrec_ablation}
\end{figure}

\begin{table}[t]
\centering
\begin{tabular}{lcc}
\toprule
\multicolumn{1}{c}{Models} & SID-1 Recall@20 (\%) & SID-1 Recall@50 (\%) \\ \midrule
X-Rec w/o AC & 28.57 & 35.77 \\
X-Rec & \textbf{41.37} & \textbf{53.55} \\ \bottomrule
\end{tabular}
\caption{Average Recall@20 and Recall@50 of the first SID digit across all streaming partitions. Recall is defined as the fraction of test examples for which the first SID digit of the ground-truth item is covered by the first SID digits of the generated triggers after quantization.}
\label{tab:sid-1_hitrate}
\end{table}

We implement an X-Rec variant without anchor conditioning, denoted as X-Rec w/o AC. In this variant, the model directly generates the recall triggers without the guidance of anchors, and all other settings remain identical to those of X-Rec. As shown in \cref{fig:xrec_ablation}, anchor conditioning plays a crucial role in X-Rec. Removing it reduces the averaged \(\mathrm{Recall}@20\) from \(10.76\%\) to \(8.71\%\), i.e., a dropping of \(2.05\%\) percentage points. Meanwhile, X-Rec w/o AC is inferior to X-Rec on all 47 streaming partitions.

To further understand the role of anchor conditioning, we quantize the generated triggers of X-Rec and X-Rec w/o AC back into SIDs, and examine whether the first SID digit matches that of the ground-truth item. As shown in \cref{tab:sid-1_hitrate}, X-Rec achieves substantially higher first-digit recall than X-Rec w/o AC. Specifically, the recall improves from \(28.57\%\) to \(41.37\%\) at top-20 and from \(35.77\%\) to \(53.55\%\) at top-50, corresponding to absolute gains of \(12.80\%\) and \(17.78\%\) percentage points, respectively. These results indicate that anchor conditioning substantially improves the model's ability to identify the correct coarse semantic region, which in turn leads to better end-to-end retrieval performance.

Overall, anchor conditioning effectively decomposes trigger generation into coarse semantic localization followed by fine-grained refinement. This decomposition simplifies the generation task and consistently improves retrieval performance, while introducing negligible additional computational cost.

\subsubsection{Effect of RFM}\label{sec:offline-euclidean-ablation}
We implement a variant of X-Rec without RFM, denoted as X-Rec w/o RFM, by replacing RFM with rectified FM in Euclidean space~\citep{liu2022flow}, while keeping all other settings unchanged. As shown in \cref{fig:xrec_ablation}, the w/o RFM variant reaches an average \(\mathrm{Recall}@20\) of \(8.98\%\), trailing X-Rec by \(1.78\%\) percentage points. X-Rec achieves higher recall on all 47 streaming partitions, indicating that the geometry-aware flow design consistently improves trigger quality.

This improvement comes from explicitly grounding the generation process on the normalized item embedding manifold. Since item embeddings are \(\ell_2\)-normalized, valid retrieval triggers should lie on the unit hypersphere. RFM incorporates this constraint into the generative dynamics by transporting samples along spherical geodesics and constraining the velocity to the tangent space at each state. As a result, the model can devote its full capacity to learning meaningful movement within the item manifold. In contrast, under Euclidean FM, the velocity field must implicitly handle two coupled objectives: moving toward the target item and correcting deviations from the hypersphere. This additional burden makes the learned dynamics less efficient and less aligned with the normalized retrieval space, leading to less accurate retrieval triggers.

\subsubsection{Effect of Late-Interaction}\label{sec:effect-late-interaction}\label{sec:offline-generation-efficiency}

X-Rec uses late interaction to restrict each velocity-field evaluation to the last Transformer layer, i.e., \(L'=1\). Here, we study the performance--efficiency trade-off by varying the number of velocity evaluation layers \(L' \in \{1,7,14,28\}\). For all configurations, the number of denoising steps is fixed to 30. We also include X-Rec-AR as a reference. Retrieval performance is measured by the average \(\operatorname{Recall}@20{\times}1\) across streaming partitions, while efficiency is measured by the maximum observed generation throughput on a single production-grade GPU in BF16. When benchmarking throughput, each request contains a condition sequence of length 199 and requires the model to generate 20 triggers.

\begin{figure}[H]
    \centering
    \includegraphics[width=0.92\textwidth]{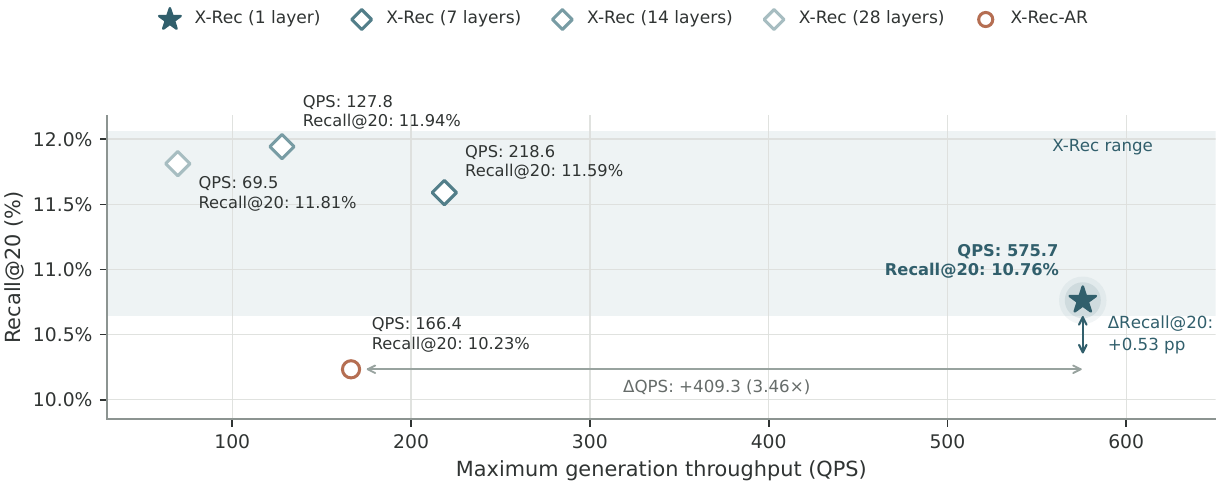}
    \caption{Performance--throughput trade-off of different models. Each point reports average Recall@20 over all streaming partitions and the maximum generation throughput across the tested batch sizes.}
    \label{fig:xrec-quality-throughput-tradeoff}
\end{figure}

As shown in \cref{fig:xrec-quality-throughput-tradeoff}, increasing \(L'\) initially improves retrieval performance, with the 14-layer variant achieving the highest mean \(\operatorname{Recall}@20\) of \(11.94\%\), which is only \(1.05\%\) percentage points higher than that of the 1-layer variant. Further increasing \(L'\) to 28, however, leads to a slight performance decline, indicating that using more layers for denoising does not necessarily improve retrieval performance. In contrast, the 1-layer variant is substantially more efficient, achieving \(2.63\times\), \(4.50\times\), and \(8.28\times\) the throughput of the 7-, 14-, and 28-layer variants, respectively. Furthermore, the 1-layer X-Rec outperforms X-Rec-AR in both retrieval performance and generation efficiency, improving mean \(\operatorname{Recall}@20\) by \(0.53\) percentage points and throughput by \(3.46\times\).

These results show that the one-layer late-interaction design offers the best deployment trade-off: it preserves strong retrieval performance while greatly reducing the repeated Transformer computation during generation.

\subsubsection{Effect of Pretraining}
To isolate the contribution of pretraining, we compare X-Rec with an otherwise identical variant trained on the streaming benchmark from scratch, i.e., without pretraining. As shown in \cref{fig:xrec_ablation}, removing the pretraining stage causes a substantial and persistent performance degradation across all the 47 streaming partitions. The mean \(\operatorname{Recall}@20\) decreases from 10.76\% to 5.49\%, an absolute drop of 5.27\% percentage points. Although the non-pretrained variant gradually improves during streaming SFT, it remains consistently underperforms the pretrained model. These results demonstrate that pretraining provides a strong and transferable initialization that cannot be fully recovered through downstream streaming supervision alone, highlighting its critical role in learning robust representations for effective retrieval.

\subsection{Analysis}\label{sec:offline-analysis-experiments}

We further analyze X-Rec in terms of trigger breadth, diversity, and the representation loss introduced by SID reconstruction. 

\subsubsection{Retrieval Breadth \& Depth}\label{sec:offline-trigger-depth-scaling}

Under a fixed candidate budget, increasing \(K\) distributes the budget across more generated triggers, broadening the coverage of user interests, whereas increasing \(R\) retrieves more nearest neighbors per trigger, enabling deeper exploration around each interest. To study this breadth--depth trade-off, we evaluate X-Rec under four \(\operatorname{Recall}@20\) configurations: \(1{\times}20\), \(5{\times}4\), \(10{\times}2\), and \(20{\times}1\). We also compare X-Rec with ComiRec-SA~\citep{cen2020controllable}, a multi-interest U2I baseline that uses self-attention to derive multiple user-interest embeddings. For a fair comparison, ComiRec-SA equips the same context-encoder configuration as X-Rec, therefore we denote its resulting variants as X-Rec-U2I-\(K\) (See \cref{app:xrec-variant-designs} for detailed implementations). We use the average pairwise cosine similarity among the generated triggers as a proxy for interest breadth, with a lower similarity indicating broader coverage. For configurations with \(K=1\), we define this pairwise cosine as 1.

\begin{figure}[t]
    \centering
    \includegraphics[width=\textwidth]{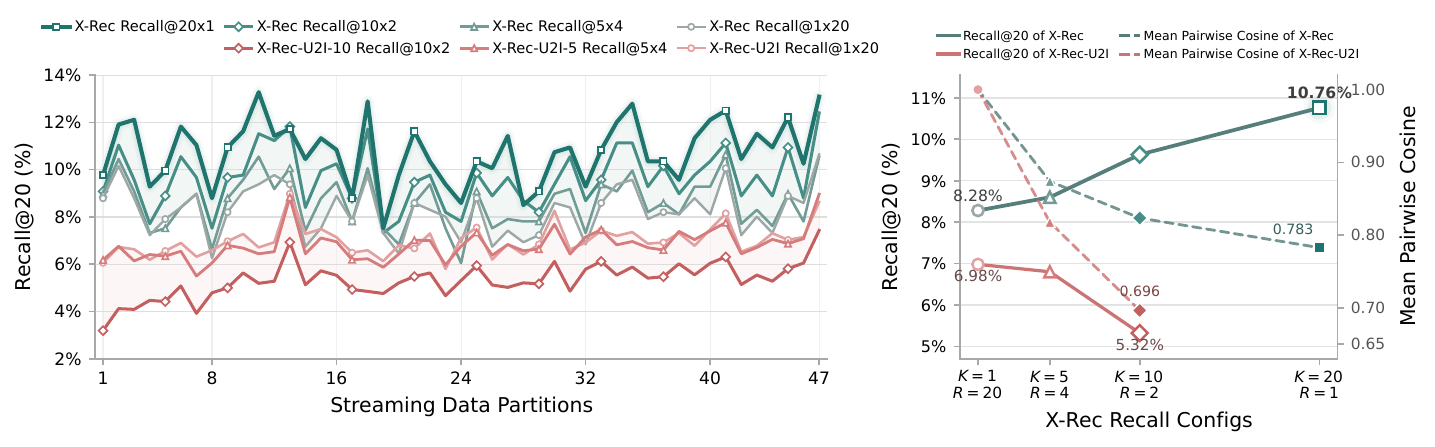}
    \caption{Performance of different recall configurations of X-Rec and X-Rec-U2I-\(K\).}
    \label{fig:xrec-trigger-depth-tradeoff}
\end{figure}

As shown in \cref{fig:xrec-trigger-depth-tradeoff}, for X-Rec, \(\operatorname{Recall}@20\) improves monotonically with the number of generated triggers \(K\). The average \(\operatorname{Recall}@20\) increases from \(8.28\%\) under \(1{\times}20\) to \(8.60\%\), \(9.63\%\), and \(10.76\%\) for \(5{\times}4\), \(10{\times}2\), and \(20{\times}1\), respectively. Reallocating the fixed budget of 20 candidates from \(1{\times}20\) to \(20{\times}1\) yields an absolute improvement of \(2.48\%\) percentage points. This gain is consistent across the evaluation stream: \(20{\times}1\) outperforms \(1{\times}20\) on every partition and achieves the best or tied-best result on 45 of the 47 partitions. Meanwhile, the average pairwise cosine similarity decreases as \(K\) increases, indicating broader coverage of user interests. Together, these results show that, under a fixed candidate budget, broadening interest coverage with more triggers is more effective than exploring deeper neighborhoods around fewer triggers. This highlights X-Rec's ability to capture multiple modes of user interest.

X-Rec-U2I-\(K\), however, exhibits a different trend. Although increasing \(K\) lowers the average pairwise cosine similarity and appears to broaden interest coverage, its \(\operatorname{Recall}@20\) drops substantially. Specifically, the average recall decreases from \(6.98\%\) under \(1{\times}20\) to \(5.32\%\) under \(10{\times}2\), a drop of \(1.66\%\) percentage points. Moreover, under the same configuration, the ComiRec-based variants X-Rec-U2I-\(K\) perform substantially worse than X-Rec. Under \(10{\times}2\), for example, X-Rec-U2I-\(K\) has a lower average pairwise cosine similarity than X-Rec (\(0.696\) vs.\ \(0.823\)), yet achieves substantially lower recall (\(5.32\%\) vs.\ \(9.63\%\)). This indicates that the apparent breadth of X-Rec-U2I-\(K\) is largely superficial: its interest embeddings are more dispersed but fail to effectively cover distinct user interests. This is because, although X-Rec-U2I-\(K\) generates multiple interest embeddings, it remains within the U2I contrastive-learning paradigm rather than explicitly modeling the full target distribution. Moreover, only the interest embedding selected for a given target is optimized during training. As a result, when \(K\) increases, the supervision received by each embedding becomes sparser, leading to performance degradation.

\subsubsection{Adjusting Diversity with Anchor Temperature}
\begin{figure}[t]
    \centering
    \includegraphics[width=\textwidth]{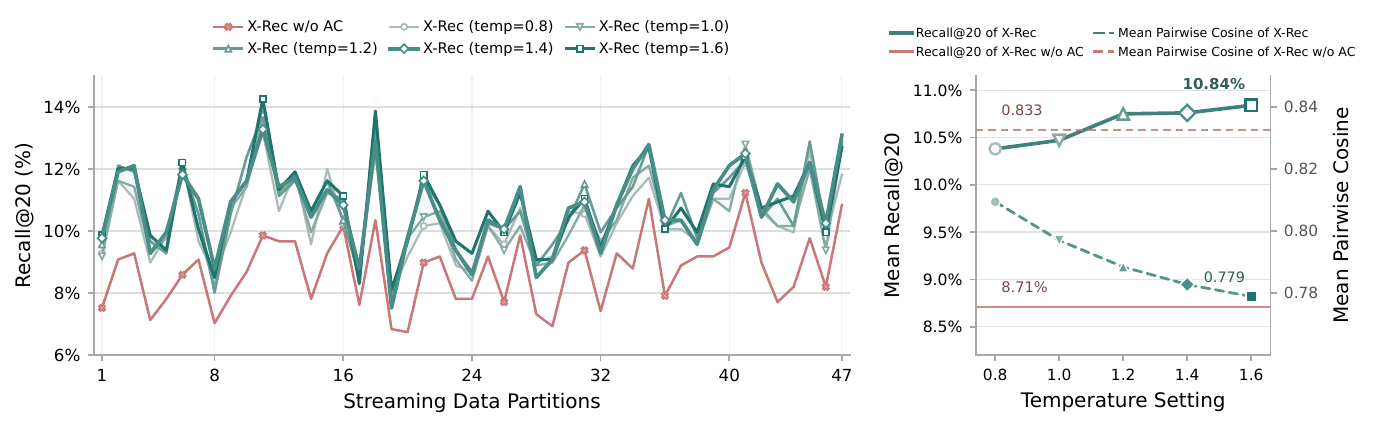}
    \caption{Effect of anchor-sampling temperature across all streaming partitions.}
    \label{fig:xrec_ablation_temperature}
\end{figure}

Beyond simplifying the generation task, anchor conditioning provides a natural mechanism for controlling trigger diversity through the sampling temperature of the anchor distribution \(p_\theta(x^{(a)} \mid c)\). As the temperature approaches zero, the anchor distribution degenerates into a delta distribution on the most probable anchor. In this case, all generated triggers are concentrated ina narrow semantic region, resulting in low diversity. As the temperature approaches infinity, the anchor distribution converges uniform over all candidate anchors. The generated triggers are then dispersed across different regions, leading to maximal diversity. In this section, we vary the temperature to study the relationship between trigger diversity and recall performance.

As shown in \cref{fig:xrec_ablation_temperature}, anchor conditioning improves both retrieval performance and trigger diversity. Even at a temperature of \(0.8\), compared with X-Rec w/o AC, X-Rec reduces the mean pairwise cosine similarity from \(0.833\) to \(0.809\), while increasing the mean \(\operatorname{Recall}@20\) from \(8.71\%\) to \(10.38\%\). Increasing the temperature from \(0.8\) to \(1.6\) further decreases the cosine similarity monotonically from \(0.809\) to \(0.779\), coupled with an increase in \(\operatorname{Recall}@20\) from \(10.38\%\) to \(10.84\%\). These results suggest that flattening the anchor distribution encourages the model to explore a broader range of semantic regions, thereby reducing redundancy among generated triggers and improving the coverage of relevant items.

Most recall improvement occurs when increasing the temperature from \(0.8\) to \(1.2\), after which the gain gradually saturates. In particular, temperatures of \(1.2\), \(1.4\), and \(1.6\) achieve \(\operatorname{Recall}@20\) values of \(10.75\%\), \(10.76\%\), and \(10.84\%\), respectively. Thus, the default temperature of \(1.4\) provides a near-saturated operating point. Overall, anchor temperature provides an effective mechanism for controlling trigger diversity and balancing concentrated exploitation against broader retrieval coverage.

\subsubsection{Reconstructed Targets Expose Representation Loss}\label{sec:offline-reconstruction-target-analysis}

Here, we examine the performance degradation caused by SID quantization errors. Instead of supervising X-Rec with the original continuous item embeddings, we construct a variant that uses SID-reconstructed embeddings as training targets. Specifically, each target item is first mapped to its SID, and the SID is then decoded through the RQ codebooks to obtain the reconstructed training target for SFT.

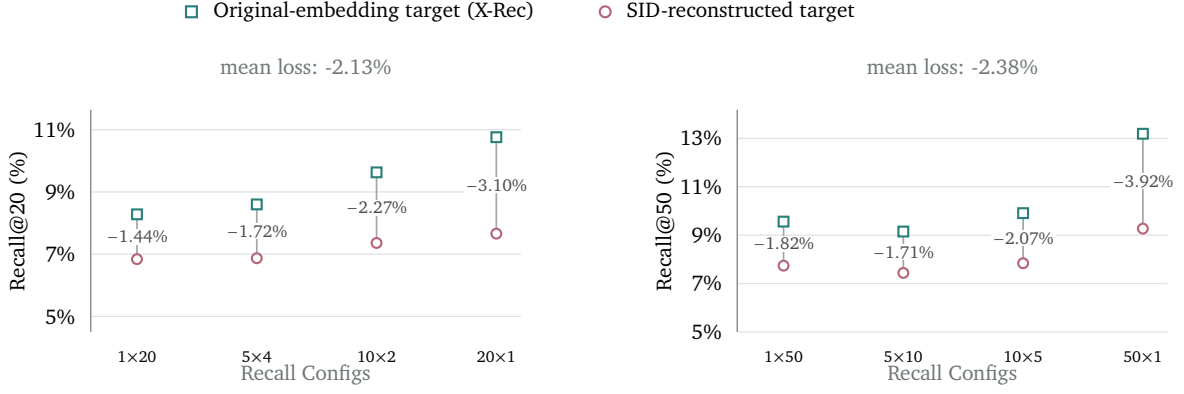
\begin{figure}[t]
    \centering
    \begin{tikzpicture}[x=0.72cm,y=0.72cm,font=\scriptsize]
        \definecolor{xrecteal}{HTML}{2A8079}
        \definecolor{sidrose}{HTML}{B56576}
        \definecolor{plotgray}{HTML}{6E7775}
        \draw[xrecteal,fill=white,line width=0.8pt] (1.30,5.78) rectangle (1.50,5.98);
        \node[anchor=west] at (1.62,5.88) {Original-embedding target (X-Rec)};
        \draw[sidrose,fill=white,line width=0.8pt] (9.00,5.88) circle[radius=0.10];
        \node[anchor=west] at (9.20,5.88) {SID-reconstructed target};
        \begin{scope}
            \node[text=plotgray] at (3.50,4.88) {mean loss: -2.13\%};
            \foreach \y/\lab in {0.286/5,1.429/7,2.571/9,3.714/11}{
                \draw[black!12,line width=0.45pt] (-0.45,\y) -- (7.45,\y);
                \node[anchor=east] at (-0.55,\y) {\lab\%};
            }
            \draw[black!45,line width=0.55pt] (-0.45,0.00) -- (-0.45,4.08);
            \foreach \x/\orig/\sid/\loss/\lab in {0.40/8.28/6.84/1.44\%/{1$\times$20},2.60/8.60/6.87/1.72\%/{5$\times$4},4.80/9.63/7.36/2.27\%/{10$\times$2},7.00/10.76/7.66/3.10\%/{20$\times$1}}{
                \pgfmathsetmacro{\yorig}{(\orig-4.5)/1.75}
                \pgfmathsetmacro{\ysid}{(\sid-4.5)/1.75}
                \pgfmathsetmacro{\ymid}{(\yorig+\ysid)/2}
                \draw[black!35,line width=0.65pt] (\x,\ysid) -- (\x,\yorig);
                \draw[xrecteal,fill=white,line width=0.8pt] ({\x-0.09},{\yorig-0.09}) rectangle ({\x+0.09},{\yorig+0.09});
                \draw[sidrose,fill=white,line width=0.8pt] (\x,\ysid) circle[radius=0.09];
                \node[font=\tiny,fill=white,inner sep=0.6pt,text=black!70] at (\x,\ymid) {$-\loss$};
                \node[anchor=north,font=\tiny] at (\x,-0.16) {\lab};
            }
            \node[text=plotgray] at (3.50,-0.78) {Recall Configs};
            \node[rotate=90] at (-1.78,2.05) {Recall@20 (\%)};
        \end{scope}
        \begin{scope}[xshift=8.55cm]
            \node[text=plotgray] at (3.50,4.88) {mean loss: -2.38\%};
            \foreach \y/\lab in {0.000/5,0.889/7,1.778/9,2.667/11,3.556/13}{
                \draw[black!12,line width=0.45pt] (-0.45,\y) -- (7.45,\y);
                \node[anchor=east] at (-0.55,\y) {\lab\%};
            }
            \draw[black!45,line width=0.55pt] (-0.45,0.00) -- (-0.45,4.08);
            \foreach \x/\orig/\sid/\loss/\lab in {0.40/9.56/7.74/1.82\%/{1$\times$50},2.60/9.15/7.44/1.71\%/{5$\times$10},4.80/9.91/7.84/2.07\%/{10$\times$5},7.00/13.19/9.27/3.92\%/{50$\times$1}}{
                \pgfmathsetmacro{\yorig}{(\orig-5.0)/2.25}
                \pgfmathsetmacro{\ysid}{(\sid-5.0)/2.25}
                \pgfmathsetmacro{\ymid}{(\yorig+\ysid)/2}
                \draw[black!35,line width=0.65pt] (\x,\ysid) -- (\x,\yorig);
                \draw[xrecteal,fill=white,line width=0.8pt] ({\x-0.09},{\yorig-0.09}) rectangle ({\x+0.09},{\yorig+0.09});
                \draw[sidrose,fill=white,line width=0.8pt] (\x,\ysid) circle[radius=0.09];
                \node[font=\tiny,fill=white,inner sep=0.6pt,text=black!70] at (\x,\ymid) {$-\loss$};
                \node[anchor=north,font=\tiny] at (\x,-0.16) {\lab};
            }
            \node[text=plotgray] at (3.50,-0.78) {Recall Configs};
            \node[rotate=90] at (-1.78,2.05) {Recall@50 (\%)};
        \end{scope}
    \end{tikzpicture}
    \caption{Effect of using SID-reconstructed embeddings as SFT targets under different recall configurations. Each marker reports the average performance over all the streaming partitions. Numbers inside the connectors denote absolute losses relative to the original-embedding target, in percentage points.}
    \label{fig:xrec-reconstruction-target-ablation}
\end{figure}

As shown in \cref{fig:xrec-reconstruction-target-ablation}, replacing original item embeddings with SID-reconstructed targets consistently hurts performance across different recall configurations. Concretely, the average recall drops by \(1.44\%\)--\(3.10\%\) percentage points for \(\operatorname{Recall}@20\) and by \(1.71\%\)--\(3.92\%\) points for \(\operatorname{Recall}@50\), with average gaps of \(2.13\%\) and \(2.38\%\) points, respectively. The degradation is largest for \(20{\times}1\) and \(50{\times}1\), where each generated trigger retrieves only its nearest ANN neighbor. In this setting, a small shift in the target representation can directly change the nearest-neighbor result. This suggests that SID reconstruction distorts the fine-grained neighborhoods of the original embedding space and introduces a non-negligible information loss. X-Rec avoids this reconstruction bottleneck by learning directly from continuous item-embedding targets.

\subsection{Case Study}\label{sec:offline-case-study}

\begin{figure*}[thb]
    \centering
    \begin{subfigure}[t]{0.98\textwidth}
        \centering
        \includegraphics[width=\linewidth]{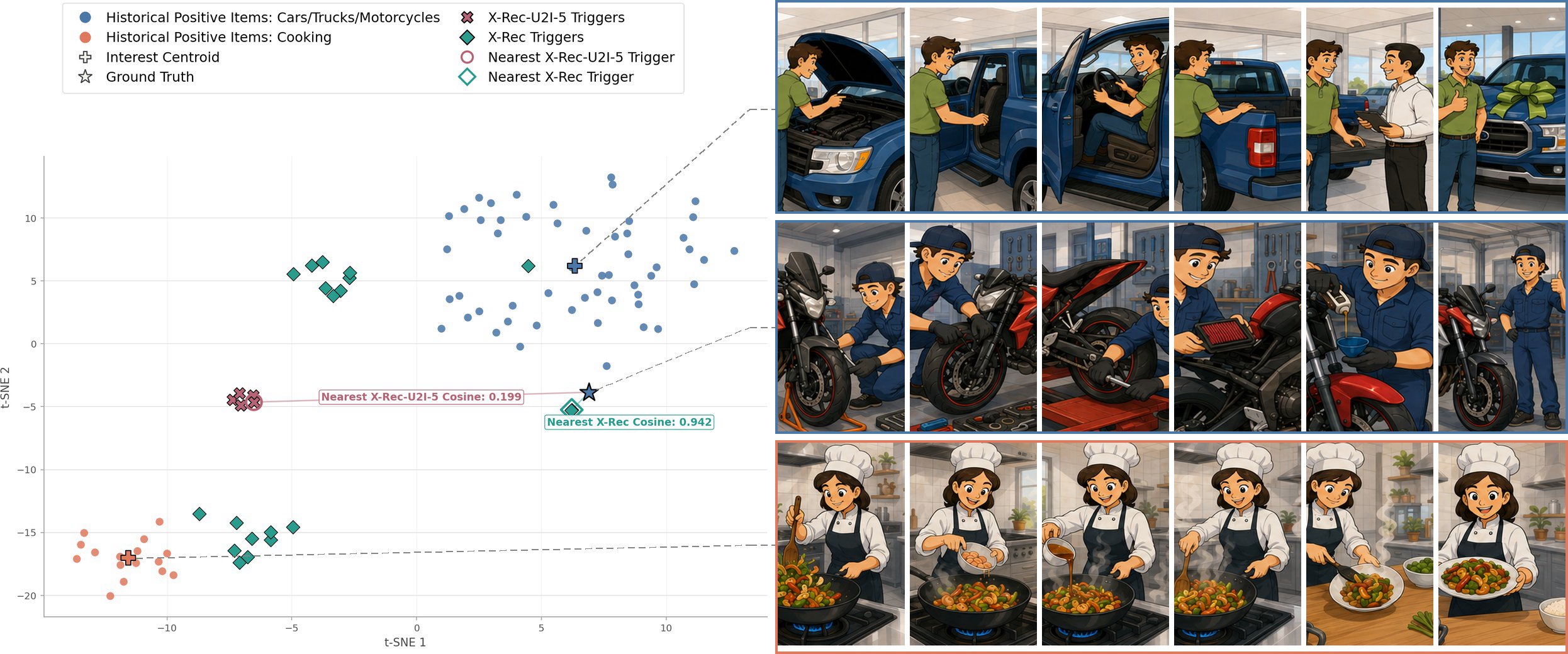}
        \caption{Active interests: Cars/Trucks/Motorcycles and Cooking. The ground-truth item belongs to Cars/Trucks/Motorcycles.}
        \label{fig:xrec-case-study-k2-cars-panel}
    \end{subfigure}
    \vspace{0.5em}
    \begin{subfigure}[t]{0.98\textwidth}
        \centering
        \includegraphics[width=\linewidth]{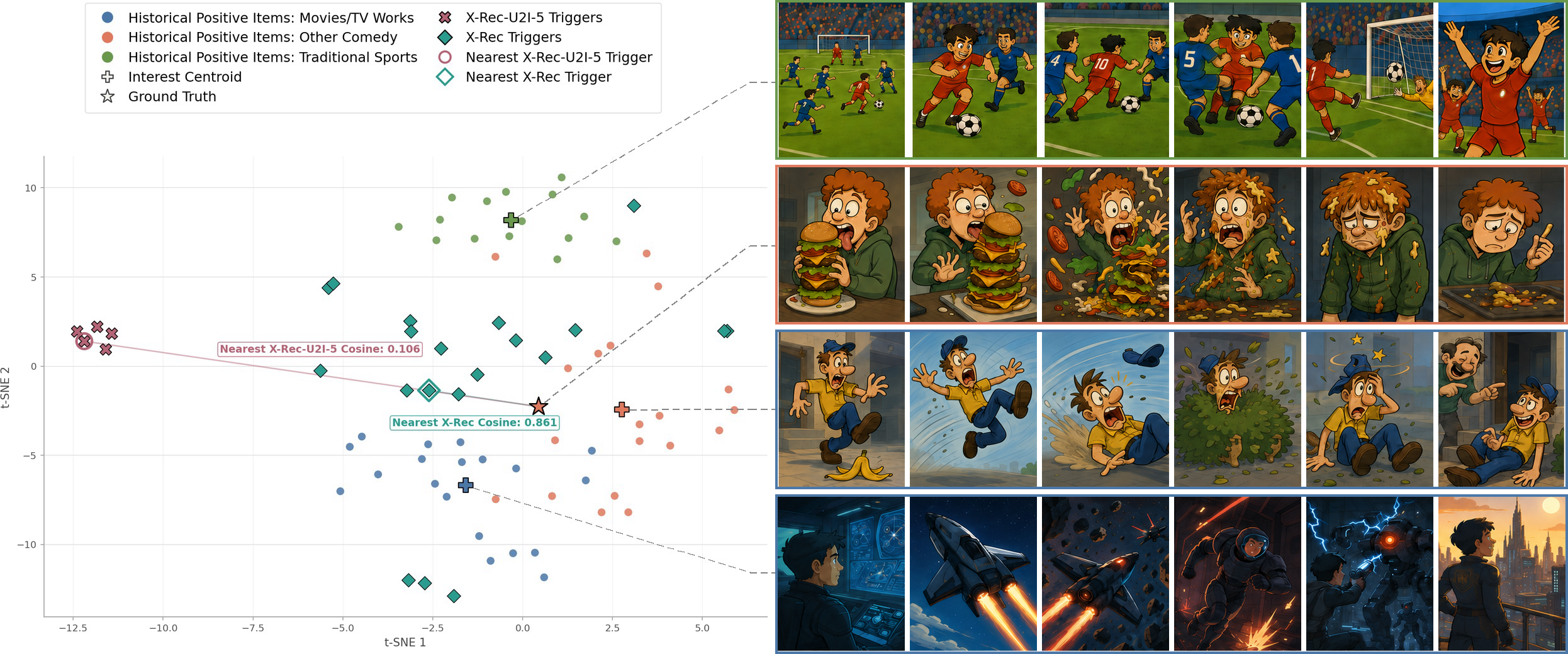}
        \caption{Active interests: Movies/TV works, Other Comedy, and Traditional Sports. The ground-truth item belongs to Other Comedy.}
        \label{fig:xrec-case-study-k3-sports}
    \end{subfigure}
    \caption{Visualization of historical positive items and triggers generated by X-Rec and X-Rec-U2I-5.}
    \label{fig:xrec-case-study-k2-cars}
\end{figure*}

In this section, we present case studies of the generated results. Specifically, we use t-SNE~\citep{cai2022theoretical} to project the 128-dimensional embeddings into two dimensions and visualize the user's historical positive items together with the triggers generated by X-Rec and X-Rec-U2I-\(K\). Representative examples\footnote{For illustrative purposes and to meet compliance requirements, all example videos are AI-generated based on the corresponding video categories and do not correspond to real user videos.} are shown in \cref{fig:xrec-case-study-k2-cars}, and additional cases are provided in \cref{app:xrec-additional-case-studies}.

As shown, user interests, represented by items with historical positive interactions, often exhibit a multi-mode structure. For example, in \cref{fig:xrec-case-study-k2-cars-panel}, the positively interacted items in the user's history mainly fall into two categories: Cars/Trucks/Motorcycles and Cooking, forming two distinct modes in the video embedding space. The user's next positively interacted ground-truth video is also likely to lie near one of these modes: in this case, it belongs to the Cars/Trucks/Motorcycles category. X-Rec closely approximates this distribution. The manifold formed by its generated triggers aligns well with that of the historical positive items, and the closest cosine similarity between the generated triggers and the ground-truth item reaches \(0.942\). In contrast, X-Rec-U2I-5 produces an averaged representation. Although it generates five distinct triggers, its manifold collapses to a single mode located in the valley between the user's two interest modes. As a result, it fails to capture the user's multi-mode interest distribution, and its closest cosine similarity to the ground-truth item is only \(0.199\). This visualization further demonstrates that, compared with U2I methods, X-Rec is able to model diverse user interests more effectively .

\FloatBarrier

\section{Online Results}\label{sec:online-results}

We deploy X-Rec as a new retrieval source for vertical-content recommendation on TikTok. In the first launch (V1, i.e., X-Rec w/o anchor conditioning), X-Rec incorporated RFM and late interaction, achieving strong generation quality with high computational efficiency. In the second launch (V2, i.e., X-Rec), we further introduced anchor conditioning, which led to additional gains in retrieval performance. As shown in \cref{tab:xrec-online-results}, X-Rec has yielded significant improvements in both vertical
engagement (+4.1484\%) and general engagement metrics (+0.0111\%). This result highlight the real-world impacts of our proposed retrieval method.

\begin{table}[htbp]
\small
\centering
\label{tab:xrec-online-results}
\begin{tabular}{lccc}
\toprule
\multicolumn{1}{c}{Metrics} & V1 & V2 & Total \\ \midrule
Vertical Engagement & +2.3779\% & +1.7705\% & +4.1484\% \\
General Engagement & Not significant & +0.0111\% & +0.0111\% \\ \bottomrule
\end{tabular}
\caption{The online A/B testing results of X-Rec. Since V1 serves as the baseline for V2 in the A/B test, the gains from the two tests can be safely combined to estimate the total improvement. Unless marked as ``Not significant'', all results in the table are statistically significant.}
\end{table}

\section{Conclusion~\&~Future Work}\label{sec:conclusion}
In this work, we presented X-Rec, an FM-based generative retrieval framework that directly models the recommendation distribution in the continuous item embedding space. Unlike U2I methods, which compress user interests into one or a few deterministic retrieval points, X-Rec generates multiple stochastic triggers to capture diverse interest modes. Unlike SID-AR methods, X-Rec avoids quantizing item embeddings into discrete identifiers and eliminates token-by-token decoding. To enable effective and efficient continuous generation at industrial scale, X-Rec incorporates three complementary designs: anchor conditioning decomposes generation into coarse semantic-region selection and fine-grained refinement; RFM aligns the transport paths with the hyperspherical geometry of normalized item embeddings; and the late-interaction DiT reuses cached condition representations while restricting repeated velocity-field estimation to the final Transformer layer.

Offline experiments on the streaming evaluation benchmark demonstrate that X-Rec achieves a favorable balance among retrieval quality, interest coverage, and generation efficiency. It outperforms both the U2I and SID-AR variants while achieving \(3.46\times\) higher generation throughput than the latter. In online experiments, two consecutive launches on TikTok's vertical content yield significant improvements on business metrics. These results demonstrate the potential of continuous generative retrieval as an expressive, efficient, and practical complement to existing industrial retrieval paradigms.

Several directions remain open for future work. First, the current sampler requires multiple denoising steps, motivating the exploration of few-step distillation, adaptive numerical solvers, and hardware-aware optimization to reduce end-to-end serving costs. Second, although the late-interaction design is computationally efficient, its scalability remains to be fully explored. Under strict serving-throughput constraints, practical systems must balance retrieval quality against generation cost. Future work could investigate how late interaction scales with larger backbones and deeper denoising modules, as well as adaptive computation mechanisms that dynamically balance quality and efficiency. Third, reward-aligned post-training, such as reinforcement learning, represents a promising direction for continuous generative retrieval. It may enable X-Rec to optimize for long-term user satisfaction and content relevance, objectives that are difficult to capture through supervised learning alone.

\printbibliography

\newpage
\quad \\
\quad \\

\section{Contributors}
Within each role, the contributors are listed in alphabetical order by their first names. Contributors marked with \textsuperscript{*} had left the team before publication.

\paragraph{Corresponding Author.}
Kun Xǔ (\url{daniel.chen28@bytedance.com})
\paragraph{Project Leader.}
Zhiwei Wang (\url{heyisen@bytedance.com})

\paragraph{Core Contributors.}
Chenglei Shen, Chenzhe Huang, Dong Jiang, Hongjie Gao, Jue Zhang, Kun Xú, Lincan Cai, Nan Zhuang, Pan Zhang, Shi Chen, Shunchi Zhang, Xiaoyu Ye, Yang Jin, Yu Zhang, Zhenwei An, Zhongtao Jiang

\paragraph{Contributors.}
Ao Xu, Chenghao Liu\textsuperscript{*}, Han Xia, Hantian Su, Haoyang Yao, Huixiu Jin, Jialin Li, Jianyuan Bo, Jie Li, Jingwei Li, Junwen Chen, Kailin Ding, Kangcheng Luo, Lin Chen, Linjie Wang, Mengshi Chen, Ming Zhang, Rengzhi Wang, Silong Yu, Song Jin\textsuperscript{*}, Tingyan Li, Weixin Wang, Xin Chen, Xuhang Xiao, Yiming Jia, Yinong Lin, Yifeng Yao, YongKang Zhang, Yu Zhang\textsuperscript{*}, Yuxiao Zhang, Zhangyi Chen, Zhao Wang, Zheng Li, Zixuan Wang

\newpage

\appendix
\setcounter{section}{0}

\renewcommand{\thesection}{A}
\renewcommand{\thesubsection}{\thesection.\arabic{subsection}}
\renewcommand{\thesubsubsection}{\thesubsection.\arabic{subsubsection}}

\section{Appendix}
\label{app:appendix}

\subsection{Complete Results on Streaming Benchmark}
\label{app:xrec-per-snapshot-recall}

\cref{tab:xrec-streaming-recall-20-snapshots} and \cref{tab:xrec-streaming-recall-50-snapshots} summarize the \(\operatorname{Recall}@20\) and \(\operatorname{Recall}@50\) results across different models. For U2I methods (SASRec, LRURec, DreamRec, and X-Rec-U2I), which produce a single retrieval embedding, we compute \(\operatorname{Recall}@K\) under the \(1\times K\) setting. For SID-based methods (TIGER, EAGER, LATTE, LLaDARec, and X-Rec-AR) and X-Rec, which support multiple retrieval triggers, we use the \(K\times 1\) setting.

\begingroup
    \centering
    \fontsize{8}{9}\selectfont
    \captionsetup[longtable]{position=bottom,aboveskip=10pt,belowskip=0pt}
    \setlength{\tabcolsep}{2.7pt}
    \renewcommand{\arraystretch}{1.02}
    \begin{longtable}{@{}c >{\columncolor{catTeal!7}}r *{9}{r}@{}}
        \toprule
        \textbf{Data Partition} & \textbf{X-Rec} &
        \textbf{X-Rec-U2I} & \textbf{X-Rec-AR} &
        \textbf{TIGER} & \textbf{LATTE} & \textbf{LLaDARec} &
        \textbf{DreamRec} & \textbf{EAGER} & \textbf{SASRec} &
        \textbf{LRURec} \\
        \midrule
        \endfirsthead
        \multicolumn{11}{c}{\tablename~\thetable\ (continued)}\\[2pt]
        \toprule
        \textbf{Data Partition} & \textbf{X-Rec} &
        \textbf{X-Rec-U2I} & \textbf{X-Rec-AR} &
        \textbf{TIGER} & \textbf{LATTE} & \textbf{LLaDARec} &
        \textbf{DreamRec} & \textbf{EAGER} & \textbf{SASRec} &
        \textbf{LRURec} \\
        \midrule
        \endhead
        \midrule
        \multicolumn{11}{r}{\textit{Continued on the next page}}\\
        \endfoot
        \bottomrule
        \caption{Complete \(\operatorname{Recall}@20\) results on the streaming benchmark. The X-Rec column is highlighted with light shading.}
        \label{tab:xrec-streaming-recall-20-snapshots}\\
        \endlastfoot
        1 & 9.7700 & 6.0645 & 9.8633 & 2.6562 & 5.9473 & 5.4004 & 4.7266 & 2.8027 & 0.9375 & 1.0938 \\
        2 & 11.9100 & 6.7285 & 10.7422 & 4.9609 & 5.8398 & 6.6309 & 4.8730 & 4.2383 & 1.4160 & 1.3281 \\
        3 & 12.1100 & 6.6309 & 10.9375 & 6.2891 & 6.7480 & 7.0117 & 3.9844 & 5.7227 & 0.7129 & 0.9082 \\
        4 & 9.2800 & 6.1914 & 9.1797 & 4.5703 & 6.0449 & 6.5430 & 4.2871 & 3.4766 & 0.8691 & 0.9961 \\
        5 & 9.9600 & 6.5625 & 9.6680 & 4.8730 & 5.8496 & 6.2695 & 3.9844 & 3.9551 & 0.9375 & 0.9961 \\
        6 & 11.8200 & 6.9043 & 11.9141 & 6.5332 & 4.1992 & 7.2656 & 5.3320 & 5.5176 & 1.1328 & 1.2402 \\
        7 & 11.0400 & 6.3184 & 10.9375 & 6.2695 & 7.1094 & 6.4062 & 3.6914 & 5.3027 & 0.6250 & 0.8398 \\
        8 & 8.7900 & 6.5918 & 8.1055 & 5.3516 & 6.5723 & 6.6113 & 4.2090 & 3.8379 & 0.8496 & 0.9766 \\
        9 & 10.9400 & 6.9727 & 10.6445 & 5.5273 & 6.5527 & 6.8555 & 3.9258 & 4.5996 & 0.9082 & 1.0254 \\
        10 & 11.6200 & 7.2754 & 13.2812 & 6.9043 & 8.1641 & 6.8652 & 4.4238 & 5.5469 & 1.1230 & 1.4453 \\
        11 & 13.2800 & 6.7090 & 12.6953 & 7.5293 & 8.8965 & 8.1641 & 4.1406 & 6.6504 & 0.9180 & 1.1328 \\
        12 & 11.4300 & 6.9336 & 11.7188 & 5.7910 & 7.0898 & 7.1094 & 3.7695 & 5.3809 & 0.8691 & 1.2109 \\
        13 & 11.7200 & 8.9844 & 10.9375 & 6.9922 & 8.5547 & 8.6230 & 6.9727 & 4.8926 & 1.6309 & 1.7578 \\
        14 & 10.4500 & 7.2754 & 9.7656 & 6.0156 & 7.3242 & 7.1680 & 4.3848 & 6.1035 & 0.9375 & 1.0645 \\
        15 & 11.3300 & 7.4902 & 11.3281 & 6.7969 & 8.3594 & 7.2754 & 4.7852 & 6.4746 & 1.4258 & 1.4355 \\
        16 & 10.8400 & 7.1387 & 10.3516 & 5.6348 & 6.6992 & 6.5039 & 4.4434 & 4.5020 & 1.1816 & 1.4844 \\
        17 & 8.7900 & 6.4941 & 8.3984 & 5.3320 & 6.2695 & 5.9570 & 3.6621 & 4.2188 & 0.8594 & 1.1230 \\
        18 & 12.8900 & 6.5918 & 11.0352 & 5.4688 & 6.9629 & 6.9531 & 4.1895 & 5.4883 & 0.9180 & 1.1133 \\
        19 & 7.5200 & 6.1328 & 7.2266 & 5.0781 & 6.1328 & 5.9961 & 3.4668 & 4.2383 & 0.8301 & 1.1328 \\
        20 & 9.7700 & 6.7969 & 9.5703 & 5.0488 & 6.7871 & 5.9961 & 3.6914 & 4.6387 & 0.8594 & 0.8887 \\
        21 & 11.6200 & 6.6699 & 10.1562 & 6.2305 & 7.7344 & 6.3965 & 3.9941 & 5.3711 & 1.0449 & 1.1426 \\
        22 & 10.3500 & 7.3047 & 8.8867 & 6.1230 & 7.8613 & 6.9141 & 4.5215 & 6.0352 & 1.0156 & 1.1914 \\
        23 & 9.3800 & 5.8008 & 9.1797 & 4.6973 & 6.2207 & 5.5859 & 3.6230 & 3.9062 & 0.7031 & 0.8789 \\
        24 & 8.5900 & 7.1094 & 8.7891 & 5.5762 & 6.8359 & 6.0547 & 3.9258 & 4.4727 & 1.1621 & 1.2109 \\
        25 & 10.3500 & 7.5488 & 11.4258 & 5.2930 & 7.0801 & 6.8848 & 4.7852 & 5.3223 & 1.2207 & 1.3965 \\
        26 & 10.0600 & 6.2012 & 10.2539 & 4.4727 & 6.1523 & 5.9473 & 3.7012 & 3.8867 & 0.8203 & 0.9082 \\
        27 & 11.4300 & 6.8457 & 9.9609 & 4.4336 & 5.9375 & 6.1230 & 3.7109 & 4.8828 & 0.9863 & 1.1816 \\
        28 & 8.5000 & 6.4062 & 9.3750 & 3.9844 & 4.6875 & 5.8398 & 3.6230 & 3.9844 & 0.7227 & 0.8594 \\
        29 & 9.0800 & 6.8359 & 7.9102 & 4.1504 & 6.3672 & 5.9082 & 3.9746 & 4.3750 & 0.9766 & 1.1426 \\
        30 & 10.7400 & 8.2617 & 9.8633 & 4.8340 & 7.0703 & 6.9043 & 5.2637 & 4.4141 & 1.3867 & 1.4844 \\
        31 & 10.9400 & 6.6211 & 10.0586 & 4.5605 & 6.7871 & 5.7910 & 3.7891 & 5.5566 & 0.9668 & 1.2012 \\
        32 & 9.2800 & 6.9238 & 8.0078 & 3.7500 & 6.4844 & 6.2988 & 4.1797 & 4.3359 & 0.7422 & 1.0449 \\
        33 & 10.8400 & 7.4219 & 10.6445 & 5.0293 & 6.8848 & 6.6992 & 5.1855 & 4.5898 & 1.2891 & 1.6309 \\
        34 & 12.0100 & 7.1973 & 10.3516 & 5.1172 & 7.2168 & 5.9277 & 3.9160 & 5.7520 & 1.0352 & 1.3379 \\
        35 & 12.7900 & 7.3633 & 12.0117 & 4.9609 & 7.3340 & 7.2363 & 4.1992 & 6.1328 & 0.8984 & 1.0449 \\
        36 & 10.3500 & 6.8652 & 8.5938 & 4.1797 & 6.4062 & 5.8496 & 3.6426 & 4.0723 & 0.9180 & 1.0742 \\
        37 & 10.3500 & 6.9238 & 9.1797 & 4.4727 & 6.4746 & 5.7813 & 4.0039 & 5.3223 & 0.8789 & 1.1621 \\
        38 & 9.5700 & 7.3242 & 9.2773 & 5.1855 & 8.1055 & 7.4121 & 5.1172 & 5.2930 & 1.2207 & 1.3184 \\
        39 & 11.3300 & 6.7773 & 10.5469 & 4.1602 & 6.8652 & 6.3086 & 3.9648 & 4.8340 & 0.9863 & 1.2500 \\
        40 & 12.1100 & 7.4707 & 10.5469 & 5.5469 & 7.9199 & 6.3477 & 4.2188 & 5.8105 & 1.2695 & 1.4844 \\
        41 & 12.5000 & 8.1543 & 12.4023 & 5.6250 & 8.2227 & 7.3535 & 5.2148 & 6.0742 & 1.2598 & 1.2402 \\
        42 & 10.4500 & 6.5332 & 10.2539 & 4.1406 & 6.7871 & 6.0938 & 3.6133 & 5.0098 & 0.8398 & 1.0352 \\
        43 & 11.5200 & 6.7773 & 10.5469 & 4.5898 & 7.3730 & 6.2402 & 4.1406 & 4.6094 & 0.9473 & 0.9961 \\
        44 & 10.9400 & 7.3145 & 9.6680 & 4.9316 & 7.3633 & 5.7520 & 3.8184 & 5.1855 & 1.1621 & 1.2598 \\
        45 & 12.2100 & 7.0312 & 12.1094 & 5.3418 & 7.8027 & 6.0156 & 4.4043 & 6.0449 & 1.2988 & 1.2207 \\
        46 & 10.2500 & 7.1484 & 10.0586 & 4.5801 & 4.1016 & 6.4258 & 4.5215 & 4.1211 & 1.0645 & 1.2012 \\
        47 & 13.0900 & 8.6328 & 12.5977 & 5.2246 & 6.6113 & 7.2461 & 6.2695 & 4.6289 & 1.5918 & 1.9727 \\
    \end{longtable}
\endgroup

\clearpage
\begingroup
    \centering
    \fontsize{8}{9}\selectfont
    \captionsetup[longtable]{position=bottom,aboveskip=10pt,belowskip=0pt}
    \setlength{\tabcolsep}{2.7pt}
    \renewcommand{\arraystretch}{1.02}
    \begin{longtable}{@{}c >{\columncolor{catTeal!7}}r *{9}{r}@{}}
        \toprule
        \textbf{Data Partition} & \textbf{X-Rec} &
        \textbf{X-Rec-U2I} & \textbf{X-Rec-AR} &
        \textbf{TIGER} & \textbf{LATTE} & \textbf{LLaDARec} &
        \textbf{DreamRec} & \textbf{EAGER} & \textbf{SASRec} &
        \textbf{LRURec} \\
        \midrule
        \endfirsthead
        \multicolumn{11}{c}{\tablename~\thetable\ (continued)}\\[2pt]
        \toprule
        \textbf{Data Partition} & \textbf{X-Rec} &
        \textbf{X-Rec-U2I} & \textbf{X-Rec-AR} &
        \textbf{TIGER} & \textbf{LATTE} & \textbf{LLaDARec} &
        \textbf{DreamRec} & \textbf{EAGER} & \textbf{SASRec} &
        \textbf{LRURec} \\
        \midrule
        \endhead
        \midrule
        \multicolumn{11}{r}{\textit{Continued on the next page}}\\
        \endfoot
        \bottomrule
        \caption{Complete \(\operatorname{Recall}@50\) results on the streaming benchmark. The X-Rec column is highlighted with light shading.}
        \label{tab:xrec-streaming-recall-50-snapshots}\\
        \endlastfoot
        1 & 12.0020 & 8.4277 & 12.2070 & 3.4863 & 6.9824 & 6.0840 & 4.8730 & 3.6523 & 1.5820 & 1.7773 \\
        2 & 12.7344 & 9.1309 & 13.7695 & 6.4746 & 7.4316 & 7.5879 & 5.1660 & 5.8105 & 2.4316 & 2.3730 \\
        3 & 13.6523 & 9.3750 & 14.3555 & 7.9687 & 8.3984 & 8.2520 & 4.2578 & 7.8906 & 1.4941 & 1.7969 \\
        4 & 11.7578 & 8.7500 & 13.5742 & 6.5527 & 7.7148 & 7.7832 & 4.5996 & 5.1855 & 1.5039 & 1.7383 \\
        5 & 11.8164 & 9.0723 & 13.3789 & 6.9238 & 7.5293 & 7.2070 & 4.1602 & 5.4492 & 1.8945 & 2.0898 \\
        6 & 14.4629 & 9.4043 & 15.2344 & 8.3887 & 5.4688 & 8.8086 & 5.4688 & 7.5586 & 2.1777 & 2.3340 \\
        7 & 12.3242 & 8.8965 & 14.0625 & 8.2812 & 8.8867 & 7.5977 & 3.9355 & 7.3242 & 1.3574 & 1.6504 \\
        8 & 12.3535 & 9.1797 & 11.9141 & 7.1680 & 8.1934 & 7.7441 & 4.4043 & 5.6348 & 1.4648 & 1.7285 \\
        9 & 12.2461 & 9.3848 & 13.6719 & 7.3242 & 8.3301 & 8.1543 & 4.1797 & 6.3086 & 1.6113 & 1.7773 \\
        10 & 13.8672 & 9.8730 & 16.0156 & 9.0723 & 10.2246 & 8.5352 & 4.5410 & 7.8418 & 2.0605 & 2.5293 \\
        11 & 14.4043 & 9.3750 & 16.6992 & 9.9121 & 11.2012 & 9.4824 & 4.3945 & 9.2090 & 1.6797 & 2.0312 \\
        12 & 13.2812 & 9.6582 & 14.9414 & 8.3398 & 9.2578 & 8.6328 & 4.0137 & 7.5586 & 1.6211 & 2.1191 \\
        13 & 15.0195 & 11.4746 & 13.2812 & 9.1797 & 10.5469 & 10.0293 & 7.0801 & 6.7773 & 2.5488 & 2.7734 \\
        14 & 13.8281 & 10.3809 & 13.8672 & 7.8809 & 9.1016 & 8.5352 & 4.5801 & 8.5156 & 1.9434 & 2.2168 \\
        15 & 14.4727 & 10.1758 & 14.0625 & 8.8867 & 10.5371 & 8.8281 & 5.0293 & 8.7695 & 2.4219 & 2.5000 \\
        16 & 13.0762 & 9.9805 & 13.6719 & 7.7539 & 8.8574 & 7.8613 & 4.6094 & 6.8945 & 2.1973 & 2.7051 \\
        17 & 11.2988 & 9.1504 & 11.4258 & 6.9434 & 7.9883 & 7.0508 & 3.8477 & 5.8887 & 1.6504 & 1.9434 \\
        18 & 13.6719 & 9.2480 & 14.3555 & 7.5000 & 8.8574 & 8.3594 & 4.4727 & 7.4414 & 1.6895 & 2.0312 \\
        19 & 10.9180 & 8.4766 & 9.2773 & 6.7285 & 7.8711 & 7.1094 & 3.6426 & 5.9863 & 1.5723 & 1.8652 \\
        20 & 11.8066 & 9.0332 & 11.8164 & 6.9531 & 8.5352 & 7.2363 & 3.8867 & 6.6309 & 1.6406 & 1.7090 \\
        21 & 13.1250 & 9.7754 & 13.1836 & 8.2129 & 9.6875 & 7.8125 & 4.3164 & 7.8613 & 2.1582 & 2.2266 \\
        22 & 13.0078 & 9.7949 & 11.2305 & 8.1738 & 9.9414 & 8.5156 & 4.7559 & 8.3594 & 1.9531 & 2.0996 \\
        23 & 10.9668 & 8.6523 & 12.5000 & 6.4258 & 7.8027 & 6.5527 & 3.7891 & 5.5371 & 1.3965 & 1.7090 \\
        24 & 11.9043 & 9.4727 & 10.9375 & 7.3438 & 8.8965 & 7.2266 & 4.1602 & 6.2598 & 2.1777 & 2.1484 \\
        25 & 13.1348 & 10.6445 & 14.9414 & 7.2363 & 8.9453 & 7.9492 & 4.9805 & 7.3145 & 2.3047 & 2.5488 \\
        26 & 10.8301 & 8.7402 & 14.0625 & 5.9766 & 7.7441 & 6.9531 & 3.9160 & 5.4590 & 1.8164 & 1.9238 \\
        27 & 11.8652 & 9.7266 & 12.6953 & 6.2012 & 7.5977 & 7.3926 & 3.9746 & 6.6211 & 1.9043 & 2.1191 \\
        28 & 10.9863 & 8.9062 & 13.0859 & 5.3906 & 6.2988 & 6.6699 & 3.7793 & 5.5078 & 1.3672 & 1.7578 \\
        29 & 11.5137 & 9.5410 & 10.3516 & 5.6836 & 8.1250 & 6.9336 & 4.2383 & 5.8789 & 1.9336 & 2.1387 \\
        30 & 12.9199 & 11.1914 & 12.7930 & 6.6113 & 8.9062 & 7.9687 & 5.5176 & 6.5527 & 2.4316 & 2.7930 \\
        31 & 12.0898 & 9.2383 & 13.0859 & 6.1133 & 8.5938 & 6.7676 & 4.0625 & 7.4609 & 1.6406 & 2.2559 \\
        32 & 11.9531 & 9.7168 & 11.4258 & 5.2637 & 8.1836 & 7.3535 & 4.3848 & 6.0352 & 1.4062 & 1.9336 \\
        33 & 12.4805 & 10.3320 & 13.9648 & 6.5137 & 8.5547 & 7.7637 & 5.4199 & 6.1719 & 2.3145 & 2.8516 \\
        34 & 12.8223 & 10.5078 & 13.6719 & 7.0020 & 9.3945 & 7.3242 & 4.1406 & 7.9395 & 2.1973 & 2.5879 \\
        35 & 13.3301 & 10.4102 & 15.1367 & 7.2559 & 9.4727 & 8.3984 & 4.5508 & 8.1152 & 1.8457 & 2.1191 \\
        36 & 11.7969 & 9.6484 & 11.3281 & 5.7910 & 7.9687 & 7.0605 & 3.8184 & 5.7227 & 1.6309 & 1.9238 \\
        37 & 12.5488 & 9.7949 & 12.3047 & 6.1328 & 8.3008 & 7.0020 & 4.1504 & 7.2266 & 1.6602 & 2.3730 \\
        38 & 12.9492 & 9.6094 & 12.5000 & 6.7871 & 9.6973 & 8.2422 & 5.3906 & 7.1289 & 2.0801 & 2.2266 \\
        39 & 12.4805 & 9.8535 & 13.9648 & 5.7031 & 8.8672 & 7.3730 & 4.2676 & 6.6309 & 1.9141 & 2.2070 \\
        40 & 13.7695 & 10.6543 & 13.2812 & 6.9824 & 9.8340 & 7.5391 & 4.5020 & 7.9785 & 2.4609 & 2.7539 \\
        41 & 13.9453 & 10.6641 & 14.7461 & 7.5879 & 10.4688 & 8.7207 & 5.4883 & 8.2422 & 2.0996 & 2.1582 \\
        42 & 12.2461 & 9.1895 & 12.6953 & 5.7910 & 8.8672 & 7.3438 & 3.8281 & 7.0605 & 1.6895 & 1.9336 \\
        43 & 12.6172 & 9.4824 & 13.5742 & 6.2988 & 9.3359 & 7.2852 & 4.2578 & 6.4160 & 1.7578 & 2.0996 \\
        44 & 13.1348 & 10.0977 & 12.5977 & 6.6309 & 9.2383 & 7.0996 & 4.1016 & 7.4121 & 2.2363 & 2.3535 \\
        45 & 13.4863 & 9.9707 & 15.3320 & 7.1777 & 9.8242 & 7.1582 & 4.6094 & 8.1250 & 2.1875 & 2.2656 \\
        46 & 12.1289 & 9.4824 & 12.3047 & 5.9766 & 5.5566 & 7.5586 & 4.7070 & 5.9277 & 1.8555 & 2.1289 \\
        47 & 14.0039 & 11.2207 & 15.5273 & 6.7969 & 8.2617 & 8.3789 & 6.5332 & 6.3281 & 2.5098 & 3.1543 \\
    \end{longtable}
\endgroup

\subsection{Architectures of X-Rec-AR and X-Rec-U2I-K}
\label{app:xrec-variant-designs}

X-Rec, X-Rec-AR, and X-Rec-U2I-\(K\) share the same Qwen3-0.6B backbone, user context, and item-representation space. They differ only in how retrieval triggers are obtained.

\paragraph{X-Rec-AR.}
X-Rec-AR adopts an autoregressive architecture to generate the target SID. Specifically, it uses the context representation \(s^{(m)}\) as a prepended prompt and models the distribution over the target 4-token SID sequence with a causal language-modeling head. The model is trained with a cross-entropy loss. During inference, we apply beam search to generate high-probability SIDs and reconstruct the retrieval triggers using the RQ codebooks.

\paragraph{X-Rec-U2I-\(K\).}
X-Rec-U2I-\(K\) adopts the self-attentive multi-interest extraction module from ComiRec-SA~\citep{cen2020controllable} to derive \(K\) user-interest embeddings from the user context. Given the last-layer hidden representation \(s^{(h_L)}\) of the context, we introduce learnable query vectors \(q_k\in\mathbb{R}^{d_m}\), \(k=1,\cdots,K\), to extract multiple interest triggers:
\begin{equation}
    t_k = W_t\left({s^{(h_L)}}^\top
    \operatorname{softmax}\left(s^{(h_L)} q_k\right)\right)
    \in \mathbb{R}^{d_e},
\end{equation}
where \(W_t\) is a learnable projection matrix.

During training, the original ComiRec-SA implementation uses only the winning trigger, i.e., the trigger closest to the target item \(\hat{x}^{(e)}\), to construct the contrastive learning loss:
\begin{equation}
    \mathcal{L}
    =
    -\log
    \frac{
    \exp\left(\frac{t_{k^\star}^\top \hat{x}^{(e)}}{\tau}\right)
    }{
    \exp\left(\frac{t_{k^\star}^\top \hat{x}^{(e)}}{\tau}\right)
    +
    \sum_{x'\in\mathcal{X}_{\text{neg}}}
    \exp\left(\frac{t_{k^\star}^\top \hat{x}'^{(e)}}{\tau}\right)
    },
    \quad
    k^\star=\arg\max_k t_k^\top \hat{x}^{(e)} .
\end{equation}

However, we observe that this strategy is highly unstable and leads to poor performance on our streaming benchmark. We therefore adopt an improved variant. For each candidate item, including both the target item and negative items, we first select its nearest interest trigger and then compute the contrastive loss using the similarity between each item embedding and its corresponding nearest trigger. The improved loss is defined as:
\begin{equation}
    \mathcal{L}
    =
    -\log
    \frac{
    \exp\left(\frac{t_{k_{\hat{x}}}^\top \hat{x}^{(e)}}{\tau}\right)
    }{
    \exp\left(\frac{t_{k_{\hat{x}}}^\top \hat{x}^{(e)}}{\tau}\right)
    +
    \sum_{x'\in\mathcal{X}_{\text{neg}}}
    \exp\left(\frac{t_{k_{x'}}^\top \hat{x}'^{(e)}}{\tau}\right)
    },
    \quad
    k_x=\arg\max_k t_k^\top x^{(e)} .
\end{equation}

In this way, all interest triggers receive direct optimization signals, instead of only the winning trigger being updated. This substantially improves training stability and leads to much better performance on the streaming benchmark. During inference, the learned interest triggers are used to retrieve candidates from the ANN index. Therefore, X-Rec-U2I-\(K\) represents user interest diversity with a finite set of deterministic user vectors, rather than a continuous generative distribution.

\subsection{Additional Multi-Interest Case Studies}
\label{app:xrec-additional-case-studies}

\begin{figure*}[t]
    \centering
    \begin{subfigure}[t]{0.98\textwidth}
        \centering
        \includegraphics[width=\linewidth]{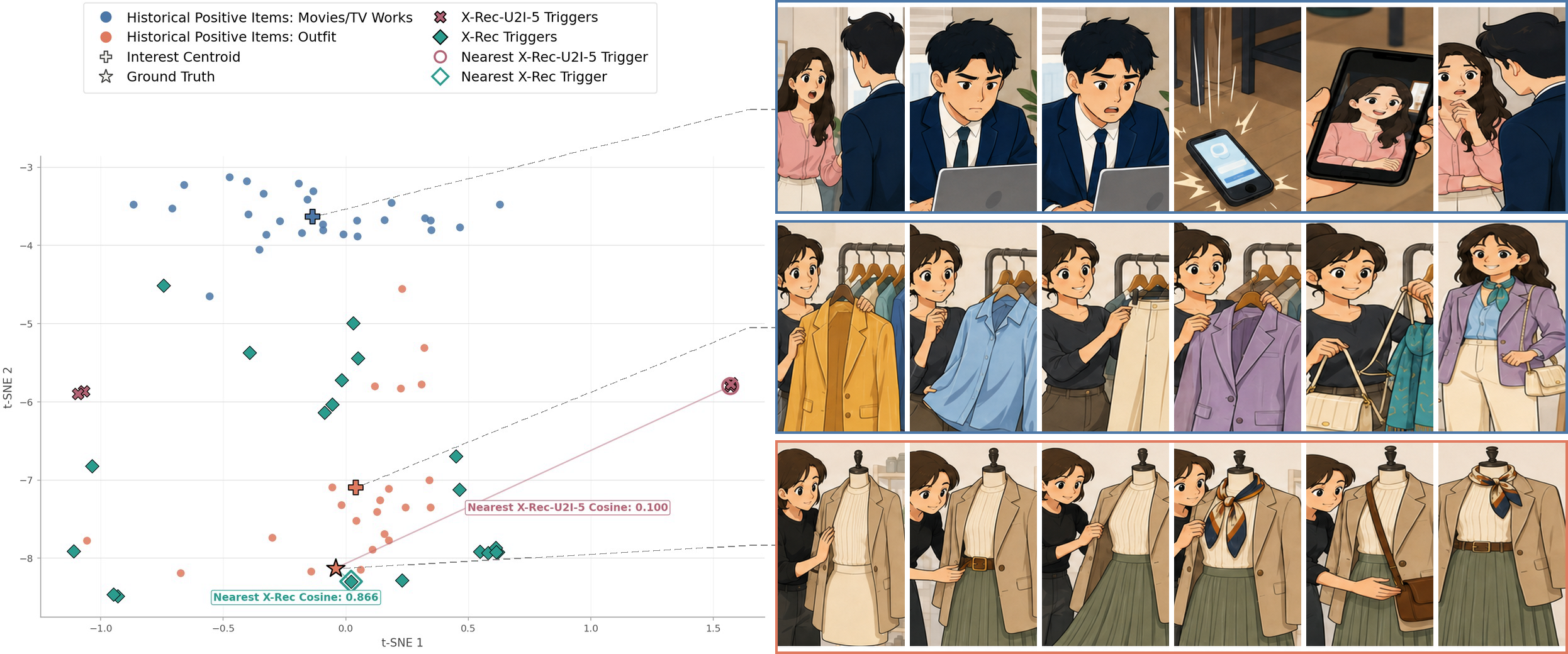}
        \caption{Active interests: Movies/TV Works and Outfit. The ground-truth item belongs to Outfit.}
        \label{fig:xrec-case-study-k2-outfit}
    \end{subfigure}
    \vspace{0.5em}
    \begin{subfigure}[t]{0.98\textwidth}
        \centering
        \includegraphics[width=\linewidth]{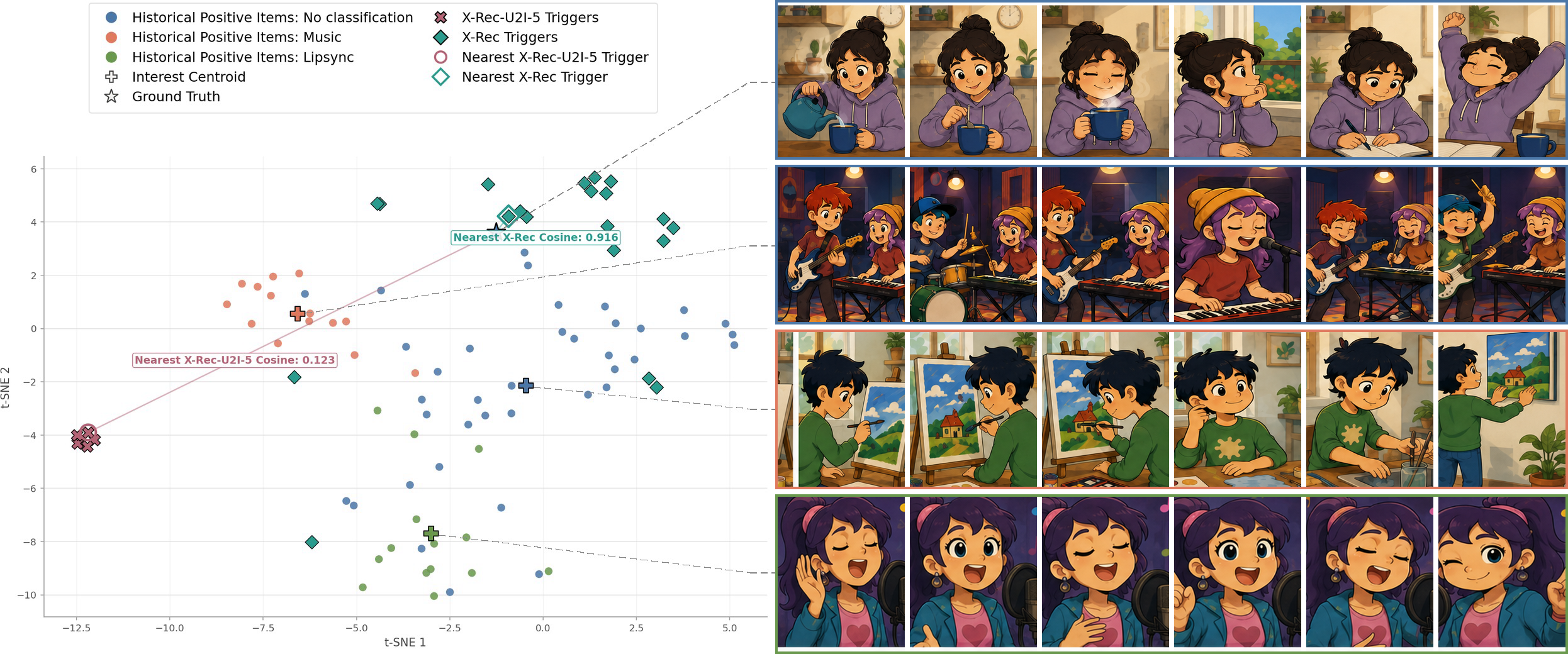}
        \caption{Active interests: Unclassified content, Music, and Lipsync. The ground-truth item belongs to the unclassified mode.}
        \label{fig:xrec-case-study-k3-unclassified}
    \end{subfigure}
    \caption{Visualization of historical positive items and triggers generated by X-Rec and X-Rec-U2I-5.}
    \label{fig:xrec-case-study-mixed}
\end{figure*}

\begin{figure*}[t]
    \centering
    \begin{subfigure}[t]{0.98\textwidth}
        \centering
        \includegraphics[width=\linewidth]{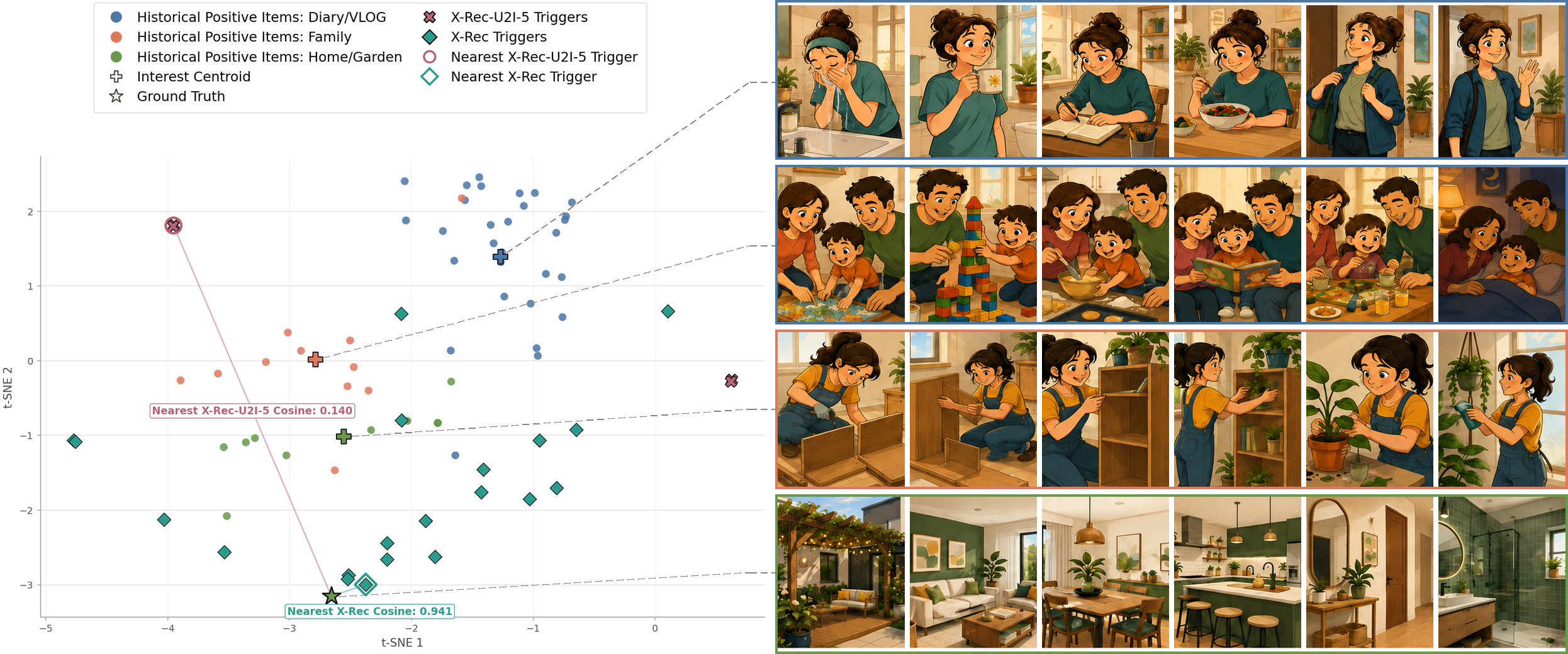}
        \caption{Active interests: Diary/VLOG, Family, and Home/Garden. The ground-truth item belongs to Home/Garden.}
    \end{subfigure}
    \vspace{0.5em}
    \begin{subfigure}[t]{0.98\textwidth}
        \centering
        \includegraphics[width=\linewidth]{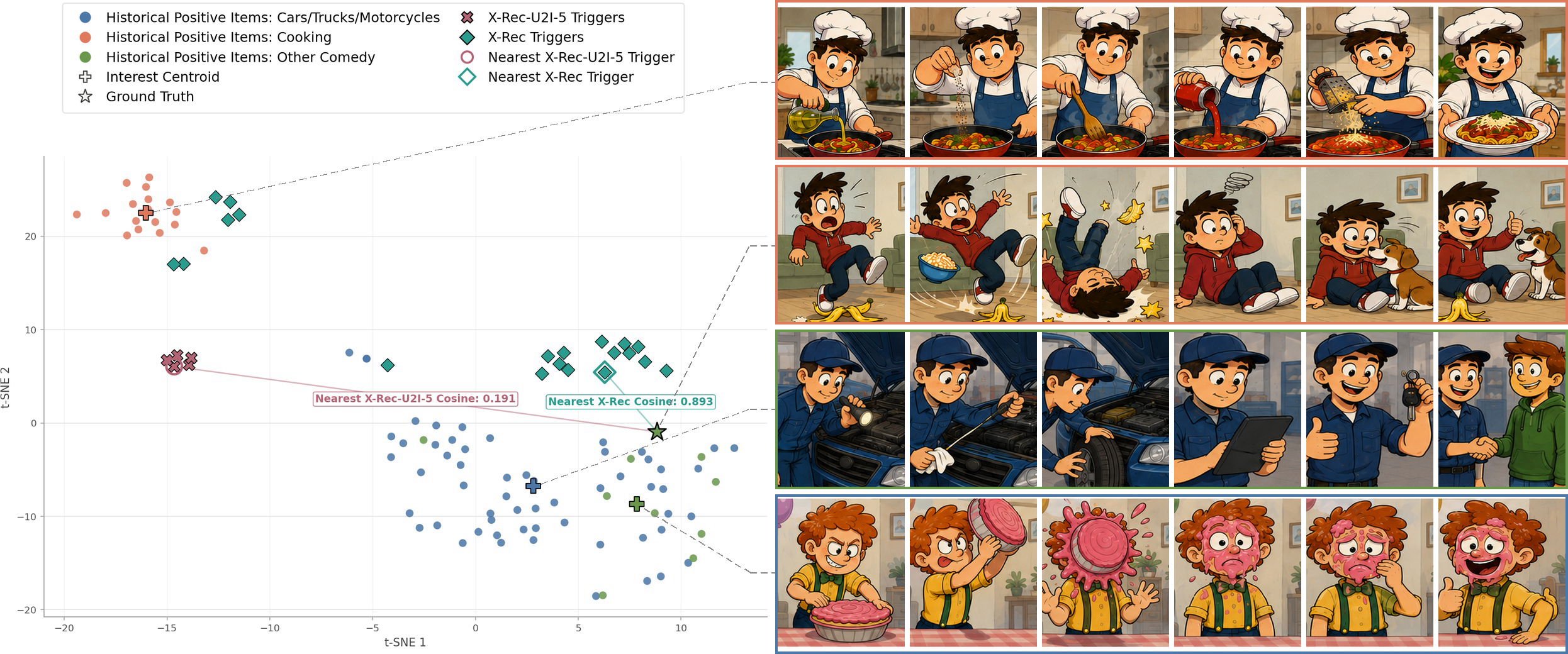}
        \caption{Active interests: Cars/Trucks/Motorcycles, Cooking, and Other Comedy. The ground-truth item belongs to Other Comedy.}
    \end{subfigure}
    \caption{Visualization of historical positive items and triggers generated by X-Rec and X-Rec-U2I-5.}
    \label{fig:xrec-case-study-k3-targets}
\end{figure*}

We provide additional qualitative examples of X-Rec-U2I-5 and X-Rec in \cref{fig:xrec-case-study-mixed} and \cref{fig:xrec-case-study-k3-targets}.

Across \cref{fig:xrec-case-study-mixed} and \cref{fig:xrec-case-study-k3-targets}, triggers of X-Rec-U2I-5 concentrate in a single local region, whereas X-Rec triggers cover multiple interest clusters. This observation supports the expressiveness claim in \cref{sec:xrec-formulation}: by directly modeling the target recommendation distribution, X-Rec can capture diverse and multi-modal user interests more effectively than U2I methods.

\begin{figure*}[t]
    \centering
    \begin{subfigure}[t]{0.98\textwidth}
        \centering
        \includegraphics[width=\linewidth]{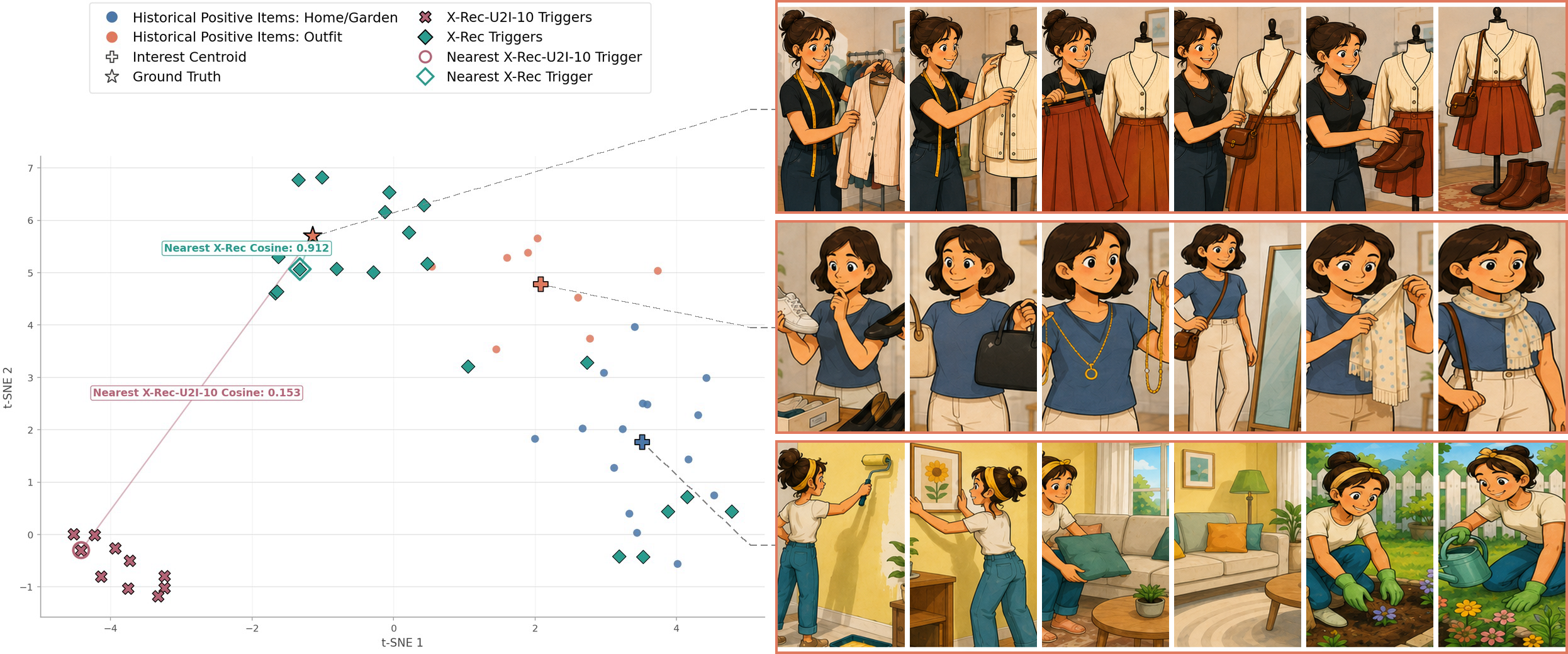}
        \caption{Active interests: Home/Garden and Outfit. The ground-truth item belongs to Outfit.}
        \label{fig:xrec-u2i-trigger10-k2-row699}
    \end{subfigure}
    \vspace{0.5em}
    \begin{subfigure}[t]{0.98\textwidth}
        \centering
        \includegraphics[width=\linewidth]{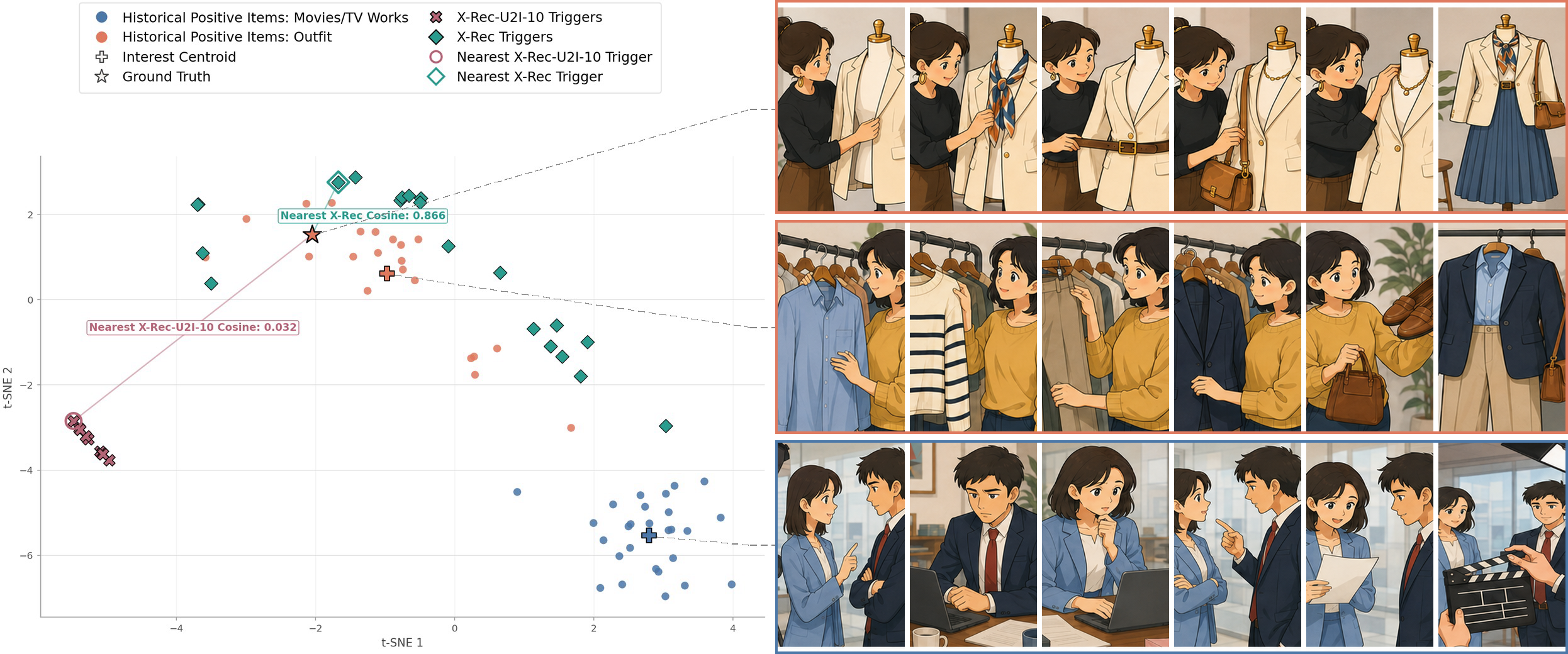}
        \caption{Active interests: Movies/TV Works and Outfit. The ground-truth item belongs to Outfit.}
        \label{fig:xrec-u2i-trigger10-k2-row502}
    \end{subfigure}
    \caption{Visualization of historical positive items and triggers generated by X-Rec and X-Rec-U2I-10.}
    \label{fig:xrec-u2i-trigger10-k2}
\end{figure*}

\begin{figure*}[t]
    \centering
    \begin{subfigure}[t]{0.98\textwidth}
        \centering
        \includegraphics[width=\linewidth]{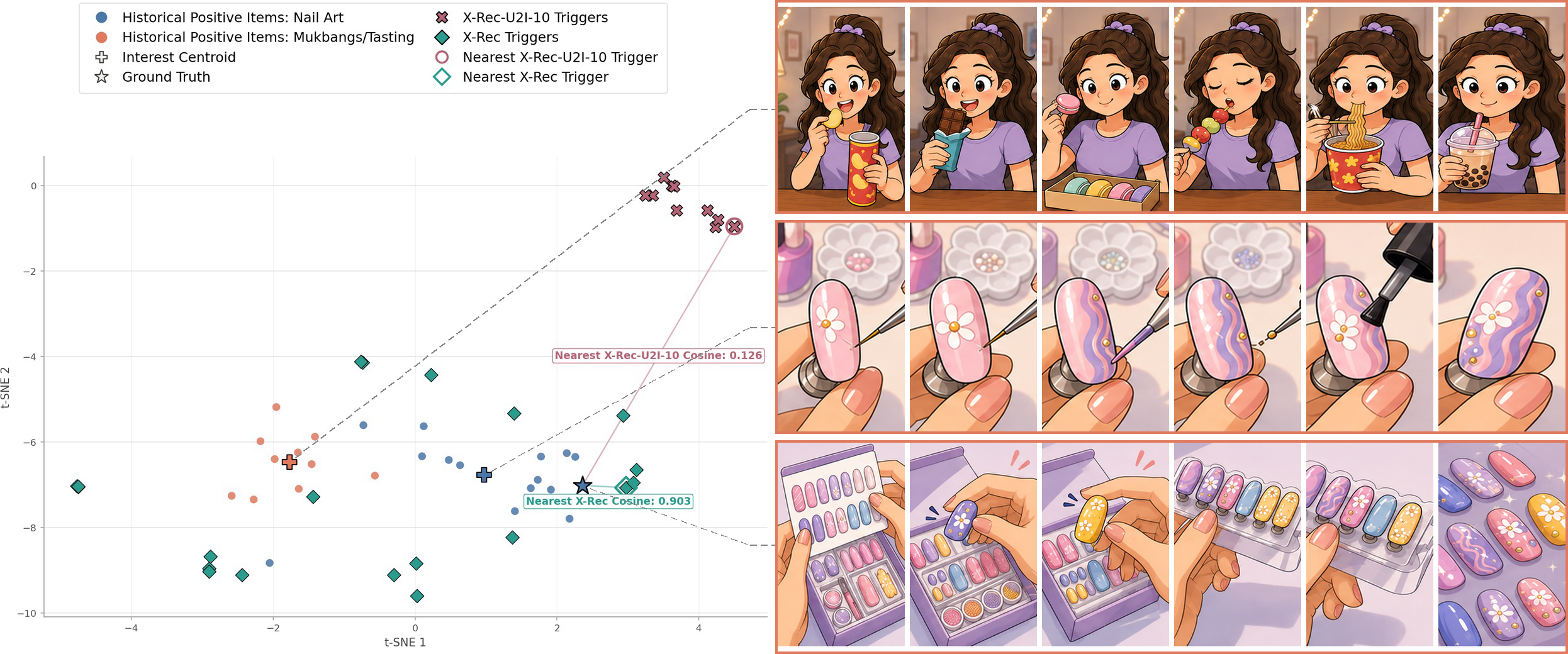}
        \caption{Active interests: Nail Art and Mukbangs/Tasting. The ground-truth item belongs to Nail Art.}
        \label{fig:xrec-u2i-trigger10-k2-row526}
    \end{subfigure}
    \vspace{0.5em}
    \begin{subfigure}[t]{0.98\textwidth}
        \centering
        \includegraphics[width=\linewidth]{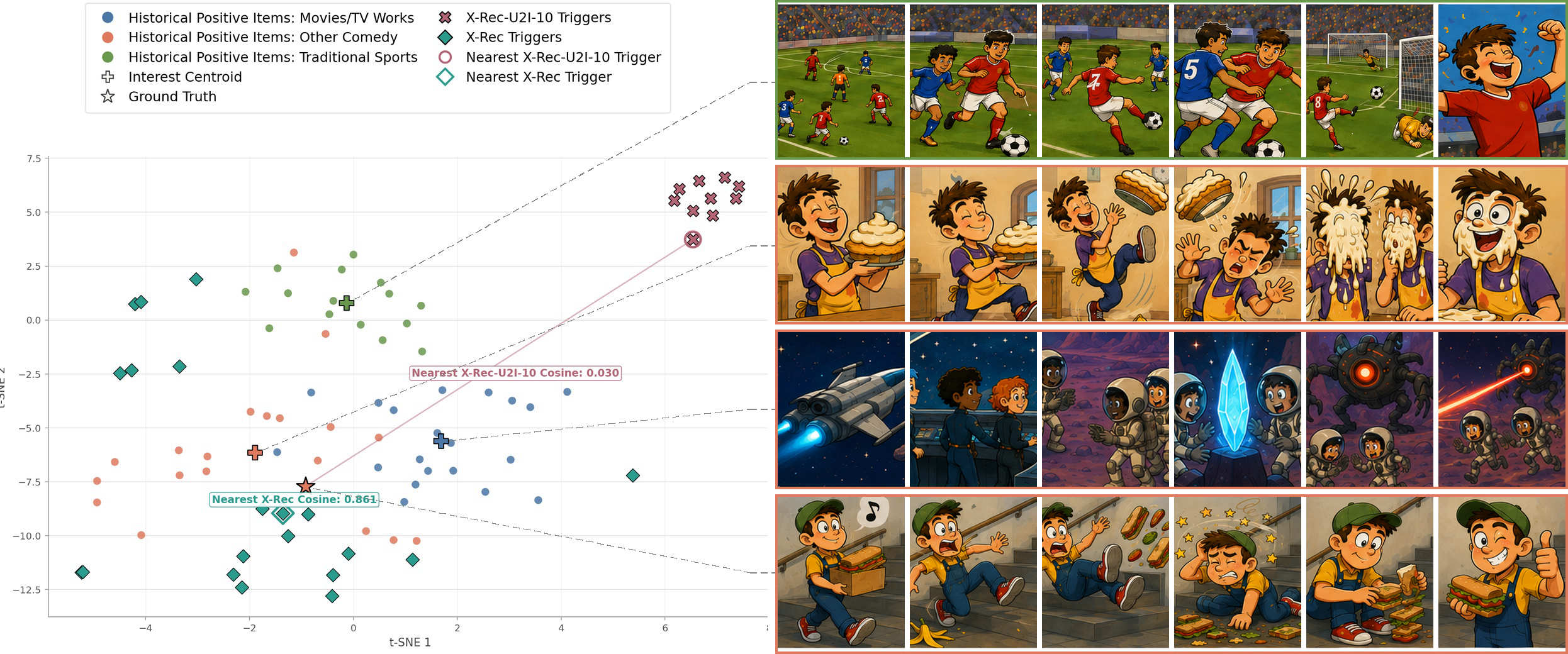}
        \caption{Active interests: Movies/TV Works, Other Comedy, and Traditional Sports. The ground-truth item belongs to Other Comedy.}
        \label{fig:xrec-u2i-trigger10-k3-row655}
    \end{subfigure}
    \caption{Visualization of historical positive items and triggers generated by X-Rec and X-Rec-U2I-10.}
    \label{fig:xrec-u2i-trigger10-mixed}
\end{figure*}

To examine whether this pattern is caused by the five-trigger budget of X-Rec-U2I-5, we increase the trigger count to 10. As shown in \cref{fig:xrec-u2i-trigger10-k2} and \cref{fig:xrec-u2i-trigger10-mixed}, increasing the trigger count does not change the concentration pattern. Therefore, simply increasing the number of U2I triggers does not eliminate the observed expressiveness gap.

\FloatBarrier

\end{document}